\documentclass[final,authoryear,1p]{elsarticle}
\usepackage{graphics}
\usepackage{epsfig}
\usepackage{bm}
\usepackage{amssymb,amsmath}
\usepackage{amsmath}
\usepackage{longtable}
\usepackage{subfig}
\usepackage{multicol}
\usepackage{supertabular}
\usepackage{bookmark}% http://ctan.org/pkg/bookmark
\usepackage{courier}
\usepackage{comment}
\usepackage{listings}
\usepackage{xcolor}
\usepackage{hyperref} %enabling hyper references
\usepackage{url} % reference by url
\hypersetup{colorlinks=true}
\begin{document}

\definecolor{mygreen}{rgb}{0,0.6,0}
\definecolor{mygray}{rgb}{0.5,0.5,0.5}
\definecolor{mymauve}{rgb}{0.58,0,0.82}
\lstset{
  backgroundcolor=\color{white},   % choose the background color; you must add \usepackage{color} or \usepackage{xcolor}
  basicstyle=\footnotesize\ttfamily,        % the size of the fonts that are used for the code
  breakatwhitespace=false,         % sets if automatic breaks should only happen at whitespace
  breaklines=true,                 % sets automatic line breaking
  captionpos=b,                    % sets the caption-position to bottom
  commentstyle=\color{mygreen},    % comment style
  deletekeywords={...},            % if you want to delete keywords from the given language
  escapeinside={\%*}{*)},          % if you want to add LaTeX within your code
  extendedchars=true,              % lets you use non-ASCII characters; for 8-bits encodings only, does not work with UTF-8
  frame=single,	                   % adds a frame around the code
  keepspaces=true,                 % keeps spaces in text, useful for keeping indentation of code (possibly needs columns=flexible)
  keywordstyle=\color{blue},       % keyword style
  language=C++,                    % the language of the code
  otherkeywords={*,...},           % if you want to add more keywords to the set
  numbers=left,                    % where to put the line-numbers; possible values are (none, left, right)
  numbersep=5pt,                   % how far the line-numbers are from the code
  numberstyle=\tiny\color{mygray}, % the style that is used for the line-numbers
  rulecolor=\color{black},         % if not set, the frame-color may be changed on line-breaks within not-black text (e.g. comments (green here))
  showspaces=false,                % show spaces everywhere adding particular underscores; it overrides 'showstringspaces'
  showstringspaces=false,          % underline spaces within strings only
  showtabs=false,                  % show tabs within strings adding particular underscores
  stepnumber=1,                    % the step between two line-numbers. If it's 1, each line will be numbered
  stringstyle=\color{mymauve},     % string literal style
  tabsize=2,	                   % sets default tabsize to 2 spaces
  title=\lstname                   % show the filename of files included with \lstinputlisting; also try caption instead of title
}

\begin{frontmatter}
\title{Direct numerical study of speed of sound in dispersed air-water two-phase flow}
%\title{This is a specimen title\tnoteref{t1}}
%\tnotetext[t1]{This document is a collaborative effort.}

\author[CSRC]{Kai Fu}
\ead{kaifu@csrc.ac.cn}
\author[CSRC,UVA]{Xiao-Long Deng\corref{cor1}}
\ead{xiaolong.deng@csrc.ac.cn}
\author[CSRC]{Lingjie Jiang}
\address[CSRC]{Beijing Computational Science Research Center, Beijing 100193, China}
\address[UVA]{Department of Mechanical and Aerospace Engineering, University of Virginia, VA 22904, USA}

\cortext[cor1]{Corresponding author.}
%\let\thefootnote\relax\footnote{Abbreviations: DNB, departure from nucleate boiling.}
%\cortext[cor1]{Corresponding author. Tel.: +46-8-5537-8887}

%Highlights:
%The interfacial area concentration transport model has been implemented in OpenFOAM.
%The Bartolomej and DEBORA experiments were selected as the validation case.
%The sensitivity of turbulent dispersion force coefficient was tested.
%The prediction of void fraction and the liquid temperature is quite satisfied.
%\input epsf

\begin{abstract}
Speed of sound is a key parameter for the compressibility effects in multiphase flow. We present a new approach to do direct numerical simulations on the speed of sound in compressible two-phase flow, based on the stratified multiphase flow model (Chang \& Liou, JCP 2007). In this method, each face is divided into gas-gas, gas-liquid, and liquid-liquid parts via reconstruction of volume fraction, and the corresponding fluxes are calculated by Riemann solvers. Viscosity and heat transfer models are included. The effects of frequency (below the natural frequency of bubbles), volume fraction, viscosity and heat transfer are investigated. With frequency 1 kHz, under viscous and isothermal conditions, the simulation results satisfy the experimental ones very well. The simulation results show that the speed of sound in air-water bubbly two-phase flow is larger when the frequency is higher. At lower frequency, for the phasic velocities, the homogeneous condition is better satisfied. Considering the phasic temperatures, during the wave propagation an isothermal bubble behavior is observed. Finally, the dispersion relation of acoustics in two-phase flow is compared with analytical results below the natural frequency. This work for the first time presents an approach to the direct numerical simulations of speed of sound and other compressibility effects in multiphase flow, which can be applied to study more complex situations, especially when it is hard to do experimental study.
\end{abstract}

\begin{keyword}
Speed of sound; Two-phase; Stratified multiphase flow method; Compressibility effects; Homogeneous flow; Bubble thermodynamics;
\end{keyword}

\end{frontmatter}

\section{Introduction}

The compressibility can lead to several physical phenomena, such as choked flow and water hammer. Consider the following equation in one-dimensional (1D) flow,
\begin{equation}\label{ru}
-M^2 \frac {{\rm d} u}{u} = \frac{{\rm d}\rho}{\rho}
\end{equation}
where $M = u/c$ is the Mach number. The derivation of Eqn. \ref{ru} can be found in \ref{1Dru}. From Eqn. \ref{ru}, we could explain the mass chocking due to compressibility effect, as discussed by \cite{Hall2015}. For low subsonic condition flow, $M$ is so small that the compressibility effects can be ignored. Thus, the increase of flow velocity means the increase of flow mass flux. As the speed of the object approaches the speed of sound $c$, the density $\rho$ starts to decrease, and the mass flux increases slower. At the critical point $M=1$, the mass flux achieves its maximum value. It is usually noted that the compressibility effects cannot be ignored when $M>0.3$.

In single-phase flow, the critical flow velocity is identical to the speed of sound of the fluid. In general, this relationship cannot apply to multiphase flow since there may be more than one speed of sound in multiphase flow. However, as discussed by \cite{Corradini2016}, the identity between the critical flow velocity and acoustic velocity of mixture is preserved in homogeneous equilibrium model (HEM). The model assumes that: (1) the velocity of each phase is equal, and (2) the phases are in thermodynamic equilibrium. Therefore, the decrease of local speed of sound can result in a decrease of local critical mass flux. The critical mass flux of the cooling system is relying on these local critical mass flux. For example in 1D case, according to the mass conservation we have $\rho u A = {\rm const}$. The critical mass flux of the system is determined by the condition (local critical mass flux) at the minimum $A$. The decrease of critical mass flux could lower the heat transfer efficiency in cooling system. Therefore, in nuclear reactor the compressibility effects can lead to a deteriorated heat transfer efficiency when the fuel assembly encounter a sudden loss of local speed of sound.

Water hammer (or hydraulic shock) is also related to the compressibility effects in the fluid. In cooling system of nuclear power plants, it can cause potential safety problem as introduced by \cite{Beuthe1997}. Water hammer, which is the generation of great variation of pressure along the mass transportation pipe, is caused by the sudden change of local fluid velocity and thus a fatigue failure may occur (see more in \cite{Calvert2000}).

Since the speed of sound is the key factor for compressibility effects, there are already extensive studies on the acoustic behavior of multiphase flow (\cite{Karplus1958,Mecredy1972,Kieffer1977,Ardron1978,Cheng1983,Ruggles1988,Costigan1997,Drew1999,Brennen2005,Simon2016,Drui2016}). In the theoretical study, an analytical expression which supposedly quantified the acoustic behavior in two-phase flow was derived from a 1D two-fluid model. Thanks to the development of high-performance computing in recent years, we are now capable to investigate acoustics of two-phase flow with a direct numerical simulation (DNS) method. In this paper, we studied the propagation of a plane wave in dispersed two-phase flow with the stratified multiphase flow method (\cite{Chang2007}). The diffuse-interface method (DIM) is applied for the direct simulation of two-phase flow with interfacial momentum and heat transfer.

The paper is structured as follows. First we introduced the theory of acoustics in homogeneous flow and separated flow. Then the governing equations for DNS study are presented, in which the viscous effects and heat diffusion are discussed. The stratified flow method adopted in the simulation is briefly introduced. The convergence of our numerical method is discussed. The momentum analysis and bubble thermodynamics are presented in the context of direct simulation results. Finally we will compare the simulated acoustic behavior in two-phase flow with both experimental and theoretical results.

\section{Theoretical speed of sound in two-phase flow}

In this section, the theory of speed of sound in two-phase flow is discussed. Consider dispersed bubbly flow in pipes, where the volume fraction is treated as constant with position along the pipe in the scale of grid size. If the size of the dispersed phase is much smaller than the wavelength, the two-phase flow resembles a single-phase fluid with effective acoustic properties as discussed by \cite{Dijk2005}. For separated flow, there are two layers clearly, with the lighter fluid flowing on top of the heavier fluid. And it has different acoustics from the bubbly flow.

In the following part, we only list the major theoretical conclusion of two-phase acoustics. The derivation could be found in \ref{dcmix}.

\subsection{Speed of sound in dispersed homogeneous flow}
\label{c_dhf}

In dispersed two-phase flow, the relative motion between dispersed and continuous phase could be ignored if the dispersed particles are much smaller than the wavelength, as indicated by \cite{Brennen2005}. We will also have a discussion in section \ref{c_momentum}. In this way, the two-phase flow could be treated as the homogeneous flow.

Consider the quiescent two-phase flow with each component uniform. We have the continuity equation,
\begin{equation}\label{alpha1}
 \frac{\partial \alpha_k \rho_k} {\partial t}+ \nabla \cdot (\alpha_k \rho_k \mathbf u_k)=0
\end{equation}

Consider the momentum conservation equation in dispersed two-phase flow,
\begin{equation}\label{NS1}
 \frac{\partial \alpha_k \rho_k \mathbf u_k} {\partial t}+ \nabla \cdot (\alpha_k \rho_k \mathbf u_k \mathbf u_k)=-\alpha_k \nabla p + \mathbf F_{i}
\end{equation}
where $\mathbf F_{i}$ refers to the interfacial momentum transfer term. Here the surface tension is ignored so that the index $k$ could be moved out from $p_k$, as shown in the first term at the right-hand side (RHS) of Eqn. \ref{NS1}. In homogenous flow, $\mathbf F_{i}$ is so large that the relative velocity is neglected.

Equations. \ref{alpha1} and \ref{NS1} are linearized and thus the speed of sound in the two-phase homogenous flow could be written as
\begin{equation}\label{chom}
\frac{1} {c_{\rm hom}^2}= (\alpha_g \rho_g + \alpha_l \rho_l)\left(\frac{\alpha_g}{\rho_g c_g^2}+ \frac{\alpha_l}{\rho_l c_l^2} \right)
\end{equation}
where
\begin{equation}\label{EOS1}
c_k^2 = \left( \frac{dp}{d\rho_k} \right)_Q
\end{equation}
Here $Q$ is the thermodynamic constraint. It is necessary to investigate the energy conservation equation to specify $Q$. In gas/liquid two-phase flow, bulk modulus of liquid is usually much larger than that of gas, that is
\begin{equation}\label{EOS1}
\rho_g c_g^2  \ll \rho_l c_l^2
\end{equation}
Thus only $c_g$ needs to be considered in Eqn. \ref{chom}. The term associated with $c_l$ can be ignored. For ideal gas, we have
\begin{equation}\label{c_iso}
c_{g}=\sqrt{\frac{\gamma_g p_g}{\rho_g}}
\end{equation}
where $\gamma_g =1.0$ for isothermal condition and $\gamma_g =1.4$ for adiabatic condition. Note that in homogeneous flow, the isothermal condition is equivalent to HEM we mentioned in the introduction. The flow has larger speed of sound in adiabatic bubble behavior, which is also discussed by \cite{Flatten2010}. We will have more discussions about this issue in section \ref{speed-thermodynamics}.

In summary, the velocity and pressure of both phases are relaxed in homogeneous flow. The speed of sound in this case could be written down as Eqn. \ref{chom}.

\subsection{Speed of sound in separated flow}

We take a 1D analysis to investigate the speed of sound in separated flow. Consider quiescent two-phase flow in a long and narrow pipe, the wave propagates along the longitude direction ($x$ axis). The wavelength is much larger than the diameter of pipe so that we have isobaric condition at each cross section of the pipe.

The continuity equation for separated flow is the same as Eqn. \ref{alpha1}. In each phase, the momentum conservation equation reads
\begin{equation}\label{NS4}
 \rho_k \left( \frac{\partial   u_k} {\partial t}+   u_k \frac{\partial u_k}{\partial x } \right )=- \frac{\partial p}{\partial x}
\end{equation}
in the longitudinal direction, where it is assumed that the flow is inviscid and there is no interfacial shear stress. The surface tension is ignored as well.

By linearizing Eqns. \ref{alpha1} and \ref{NS4}, we obtain the speed of sound in separated inviscid flow,
\begin{equation}\label{csep}
\frac{1} {c_{\rm sep}^2} \left( \frac{\alpha_g}{ \rho_g} + \frac{\alpha_l}{ \rho_l} \right)=  \frac{\alpha_g}{\rho_g c_g^2}+ \frac{\alpha_l}{\rho_l c_l^2}
\end{equation}
where $c_k$ refers to Eqn. \ref{EOS1}.

In summary, only the pressure of both phases is relaxed in separated flow. There is a relative motion between the two phases in the flow. The speed of sound in this case could be written down as Eqn. \ref{csep}.

\section{Numerical methods}
In DIM, an appropriate fluid mixture model has to be proposed to describe (1) equation of interface motion and (2) equation of state for fluid mixture in mixing region. \cite{Baer1986,Zein2010} proposed the 7-equation model for two-phase flow. The model is a full non-equilibrium model, in which each phase has its own pressure, velocity and temperature. The 6-equation model is derived from the 7-equation model in the asymptotic limit of stiff velocity relaxation, as discussed in \cite{Saurel2009}. The 7-equation or 6-equation model should not be considered as a physical model, but more as a step-model to solve the 5-equation model, which is introduced in \cite{Shyue1998, Murrone2005}. In the 5-equation model, \cite{Shyue2014} discussed that mechanical equilibrium is assumed, while the thermal and chemical relaxation are frozen. In his work, a comparison of equilibrium speed of sound in steam/water two-phase flow is made under different relaxation limits.

In some practical problems, due to the slow effects of the fluid viscosity, the velocity in tangential direction may have a large relaxation time scale. Therefore in our work, we applied the model which assumes that the velocity of each phase is non-equilibrium. In this section, we introduced the stratified multiphase flow method which was implemented in {\it Taiji} solver by \cite{Chang2007}. In this method, they studied the compressible multifluid equations in which the fluids are assumed inter-penetrating, non-homogeneous and non-equilibrium. That is to say, each fluid has its own velocity and temperature fields at the same location, but all fluids share the same pressure.

\subsection{Governing equation}
First, we introduce the governing equation used in our method. The continuity equation is already introduced as in Eqn. \ref{alpha1}. The inviscid Navier-Stokes equation was introduced as,
\begin{equation}\label{uk1}
\frac{\partial (\alpha_k \rho_k {\mathbf u}_k)}{\partial t} + \nabla \cdot (\alpha_k \rho_k {\mathbf u}_k {\mathbf u}_k)
+ \nabla(\alpha_k p_k) =  p_i \nabla \alpha_k
\end{equation}
The first and second term in the left-hand side (LHS) is the convection term. The third term is similar to the typical pressure interaction term in single-phase flow. The RHS represents the interfacial momentum transfer term and only includes the pressure contribution part.

In our method, we assumed that both components have the same pressure, thus
\begin{equation}
p=p_k
\end{equation}
Furthermore, we assumed that the interfacial pressure is equal to the pressure of each fluid, as suggested by \cite{Ishii2011} (p.187),
\begin{equation}
p=p_i
\end{equation}

If the viscosity is considered then we finally obtain the momentum conservation equation as
\begin{equation}\label{moment}
\frac{\partial (\alpha_k \rho_k {\mathbf u}_k)}{\partial t} + \nabla \cdot (\alpha_k \rho_k {\mathbf u}_k {\mathbf u}_k)
+ \nabla(\alpha_k p) - \nabla \cdot (\alpha_k {\boldsymbol \tau}) =  p \nabla \alpha_k
\end{equation}
where $\boldsymbol \tau$ is the averaged stress,
\begin{equation}
\boldsymbol \tau = \mu (\nabla \mathbf u + (\nabla \mathbf u)^{+} ) - \frac{2}{3}\mu(\nabla \cdot \mathbf u)\mathbf I
\end{equation}

\begin{equation}
\mathbf u = \sum_k \alpha_k \rho_k {\mathbf u}_k
\end{equation}

\begin{equation}
\mu = \sum_k \alpha_k \rho_k \mu_k
\end{equation}

\cite{Chang2007} introduced the energy conservation equation for inviscid two-phase flow as,

\begin{equation}\label{ek2}
\frac{\partial (\alpha_k \rho_k e_k)}{\partial t} + \nabla \cdot (\alpha_k \rho_k e_k {\mathbf u}_k)+\nabla \cdot (\alpha_k p {\mathbf u}_k) =  -p \frac{\partial \alpha_k}{\partial t}
\end{equation}
where $e_k$ refers to the virtual internal energy of phase $k$ which includes the standard thermal energy and the turbulent kinetic energy (see more in \cite{Ishii2011}). The first and second term in LHS of Eqn. \ref{ek2} is the convection term. The third term is similar to the work done by pressure in single-phase flow. The RHS represents the work done by interfacial pressure to phase $k$ due to its volume change.

Equation \ref{ek2} is applied to inviscid two-phase flow without heat conduction. Furthermore
considering the viscosity and heat conduction, we have the energy conservation equation as,
\begin{equation}\label{ek}
\frac{\partial (\alpha_k \rho_k e_k)}{\partial t} + \nabla \cdot (\alpha_k \rho_k e_k {\mathbf u}_k)+ \nabla \cdot (\alpha_k (p  {\mathbf I} - {\boldsymbol \tau}) \cdot \mathbf{u})+\nabla \cdot (\alpha_k \mathbf q^{\prime \prime}) =  -p \frac{\partial \alpha_k}{\partial t}
\end{equation}
where the average heat flux is calculated with the Fourier's law,
\begin{equation}
{\mathbf q^{\prime \prime}} =-k \nabla T
\end{equation}
Here the average conductivity of mixture is calculated as
\begin{equation}
k = \sum_k \alpha_k \rho_k k_k
\end{equation}
and the average temperature of mixture
\begin{equation}
T = \frac{\displaystyle \sum_k \alpha_k \rho_k c_{pk} T_k}{ \displaystyle \sum_k \alpha_k \rho_k c_{pk} }
\end{equation}

In Eqn. \ref{ek}, we simply assign $\alpha_k$ as the ratio that phase $k$ obtained from the conduction heat flux $\mathbf q^{\prime \prime}$. The method was not strictly developed from two-phase model and the accurate modelling of heat conduction in two-phase flow is beyond scope of this work.

It is suggested by \cite{Brennen2005} that the isothermal bubble behavior is more favorable during wave propagation. Therefore, similar to the work done by \cite{Fu2017}, we add an additional source term and rewrite Eqn. \ref{ek} as
\begin{equation}\label{ek3}
\frac{\partial (\alpha_k \rho_k e_k)}{\partial t} + \nabla \cdot (\alpha_k \rho_k e_k {\mathbf u}_k)+ \nabla \cdot (\alpha_k (p  {\mathbf I} - {\boldsymbol \tau}) \cdot \mathbf{u})+\nabla \cdot (\alpha_k \mathbf q^{\prime \prime}) =  -p \frac{\partial \alpha_k}{\partial t}+ S_k
\end{equation}
where $S_k$ is the source term to keep gas at constant temperature,
\begin{equation}
S_g = \frac{\alpha_g \rho_g}{\Delta t}c_{pg}(T_{\rm ref}-T_g)
\end{equation}
Here $T_{\rm ref}$ is the reference temperature and $\Delta t$ is numerical time step. To preserve the conservation of energy, we have
\begin{equation}
S_l = -S_g
\end{equation}

We call Eqn. \ref{ek} as thermal model 1 (TM1) and call Eqn. \ref{ek3} as thermal model 2 (TM2). Both models are studied in the simulation.

In summary, Eqns. \ref{alpha1}, \ref{moment} and \ref{ek3} are used as the governing equations in our simulation.

\subsection{Perturbation boundary condition}
\begin{figure}
\vspace*{-0.5cm}
\hspace*{3.0cm}
\includegraphics[scale=0.6]{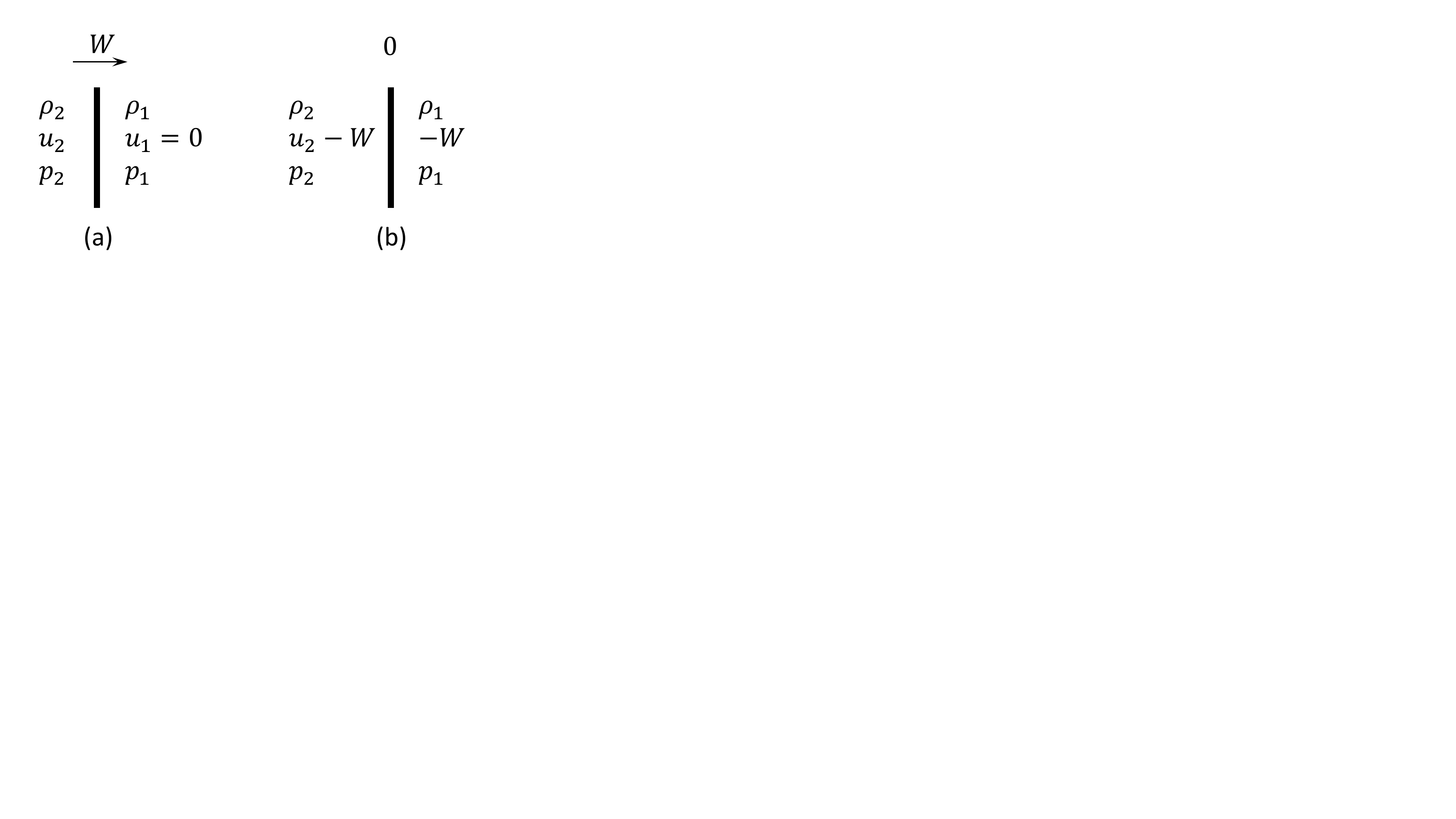}
\vspace*{-8.5cm}
\caption{Right-facing shock wave: (a) stationary frame of reference, shock has
speed $W$; (b) frame of reference moves with speed $W$, so that the shock has zero
speed}
\label{fig01}
\end{figure}

To investigate the wave propagation, we have to setup an appropriate boundary condition which represents the vibration source. Naturally we would like to setup a boundary condition with a harmonic perturbation. The perturbation condition we applied here are obtained from the Rankine-Hugoniot condition, as discussed by \cite{Toro2009} and \cite{Ranjan2011}. Consider the Rankine-Hugoniot condition for a shock. It is found convenient to transform the problem to a new frame of reference moving with the shock so that in the new frame the shock speed is zero. Figure \ref{fig01} depicts both frames of reference.

Consider the inviscid flow, which is adiabatic across the shock wave in Fig. \ref{fig01}b. The following equations read,
\begin{equation}\label{RH1}
\rho_1 W = \rho_2 (W-u_2)
\end{equation}

\begin{equation}\label{RH2}
p_1+\rho_1 W^2 = p_2 + \rho_2(W-u_2)^2
\end{equation}

\begin{equation}\label{RH3}
h_1+ \frac{1}{2} W^2  = h_2 +\frac{1}{2} (W-u_2)^2
\end{equation}
Here subscript 1 refers to the wave front and subscript 2 the post-shock. $W$ is the interface moving velocity, or wave velocity.
The stiffened-gas model used in \cite{Bai2017} is applied as equation of state (EOS) for fluids, as
\begin{equation}\label{psti}
p_k(\rho_k,T_k)= \frac{\gamma_k-1}{\gamma_k}c_{pk} \rho_k T_k-p_{\infty,k}
\end{equation}
where $c_{pk}$ is the specific heat capacity of phase $k$ at constant pressure. And,
\begin{equation}\label{hsti}
h_k(\rho_k,T_k)\equiv e_k+\frac{p_k}{\rho_k}=c_{pk} T_k
\end{equation}
The relevant parameters for both phases are listed in Table \ref{eos}.
\begin{table}[h]
\caption{Parameter for stiffened-gas model}
\begin{tabular}{cccc}
\hline
  fluid & $\gamma$ & $c_p$ (${\rm\:J/(kg \cdot K)}$)  & $p_\infty$ (Pa)  \\
  %     &           &  ${\rm\:J/(kg \cdot K)}$           & Pa \\
\hline
   water & 2.788103 & 4190.0 & $7.86253 \times 10^8$   \\
   air & 1.4 & 1008.7 &  0 \\
\hline
\end{tabular}
\label{eos}
\end{table}

Combining Eqns. \ref{RH1}, \ref{RH2}, \ref{RH3}, \ref{psti} and \ref{hsti}, we may obtain the post-shock condition $(\rho_2,u_2,p_2,h_2,T_2)$ if the equilibrium condition $(\rho_1,W,p_1,h_1,T_1)$ is given, as shown in Table \ref{equilibrium}.

\begin{table}[h]
\caption{Equilibrium conditions for both components}
\begin{tabular}{ccccc}
\hline
  fluid & $\rho_1$ (kg/m$^3$)   & $W$ (m/s)   & $p_1$ (bar) & $T_1$ (K) \\
 %       & kg/m$^3$ & m/s & bar  & K\\
\hline
   water & 998.23 & $1.482 \times 10^{3}$ & 1.013 & 293.15   \\
   air & 1.199 & $3.44263\times 10^{2}$ &  1.013   & 293.15\\
\hline
\end{tabular}
\label{equilibrium}
\end{table}

Define the Mach number
\begin{equation}
M = W/c
\end{equation}
where $c$ is the speed of sound in fluid (liquid or gas). In order to investigate the speed of sound, we introduced a weak shock wave with nearly unity Mach number $M=1.0000001$ in pure liquid water and $M=1.001$ in air. Besides, there might be considerable acoustic attenuation during wave propagation if the viscosity is considered. A stronger shock wave is introduced to capture the propagation wave, for example $M=1.000001$ in water and $M=1.01$ in air. The sound pressure level (SPL) is defined to describe the sound intensity as
\begin{equation}
%L_p = 20 \log_{10} \frac{p_2-p_1}{\sqrt{2}p_0}
L_p = 20 \log_{10} \frac{\bar{p}}{p_0}
\end{equation}
where the reference pressure $p_0=20$ $\mu$Pa. Here $\bar{p}$ is the root mean square pressure. For the perturbation boundary condition, we introduced a continuous wave function as
\begin{equation}\label{perturbationBC}
{\rm d} \phi = \Delta \phi \cos(2\pi f t)
\end{equation}
where $\Delta \phi = \phi_2 - \phi_1$ refers to the deviation amplitude from the equilibrium value. $f$ is the frequency and $t$ is time. In this case,
\begin{equation}
\bar{p} = \frac{p_2-p_1}{\sqrt{2}}
\end{equation}
The post-shock condition for both phases are calculated as in Table \ref{postshock}.

\begin{table}[h]
\caption{Post-shock conditions for both components}
\begin{tabular}{cccccc}
\hline
  fluid & $M$ & $L_p$ (dB)& $u_2$ (m/s) & $p_2$ (bar) & $T_2$ (K) \\
\hline
   water & 1.0000001 & 138.17 & $1.54855 \times 10^{-4}$ & 1.015291 & 293.15   \\
   water & 1.000001 & 158.26 & $1.564805 \times 10^{-3}$ & 1.036149 & 293.1505   \\
   air & 1.001 & 138.44 & 0.5729117 & 1.015365 &  293.3454\\
   air & 1.01 & 158.48 & 5.703604 & 1.036755 &  295.098\\
\hline
\end{tabular}
\label{postshock}
\end{table}

\subsection{Stratified flow method}
\begin{figure}
\vspace*{-0.5cm}
\hspace*{3.0cm}
\includegraphics[scale=0.6]{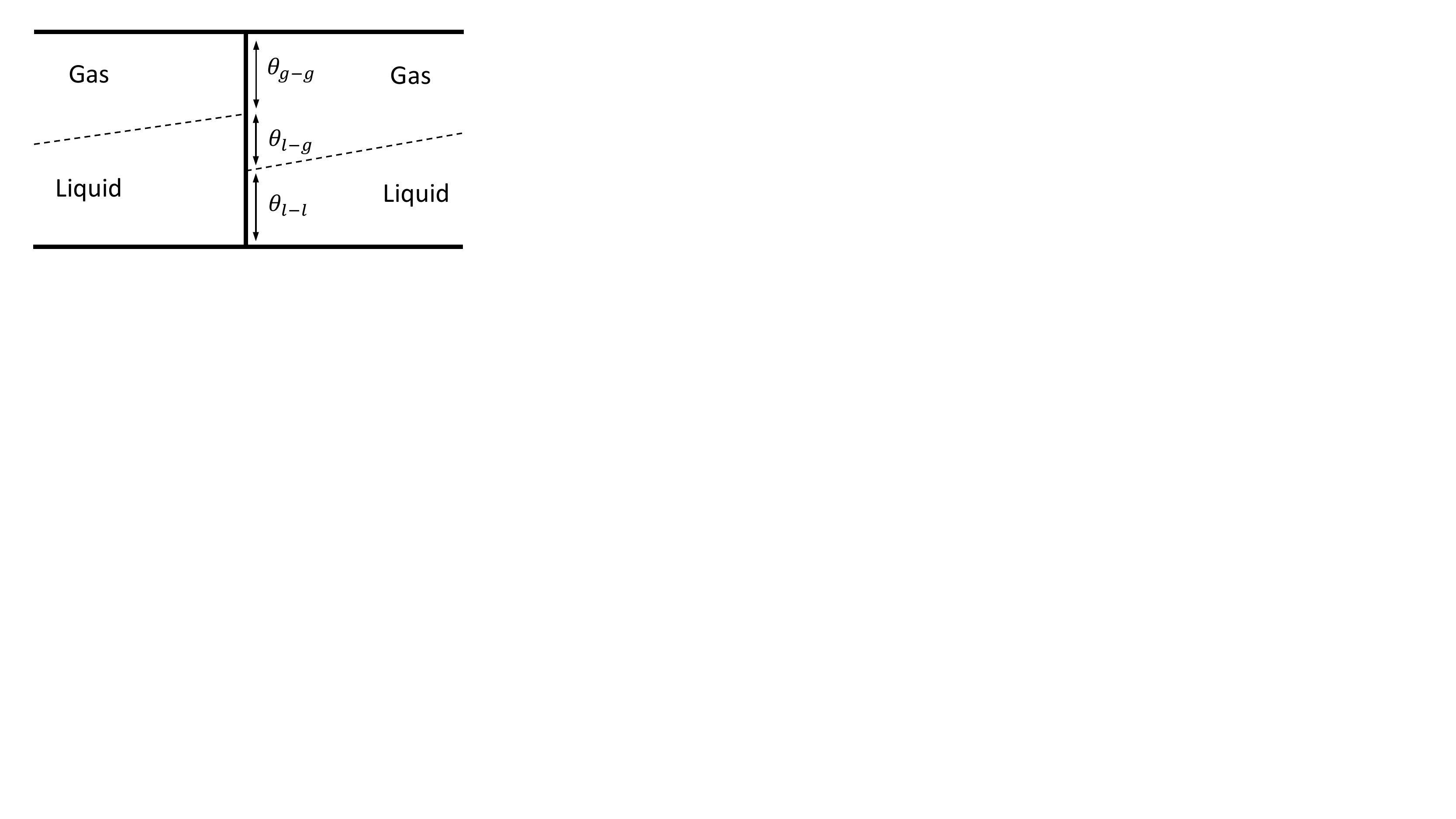}
\vspace*{-8.5cm}
\caption{Illustration of 1D stratified flow model}
\label{fig02}
\end{figure}

In this work, we applied the stratified flow method for flux calculation in finite volume method. The flow variables are described by a piecewise function within the cell. The volume fraction may be discontinuous at the cell boundaries with the reconstruction, as shown in Fig. \ref{fig02}. Therefore, the cell boundaries are naturally divided into three parts: gas-gas interface ($\theta_{g-g}$), liquid-gas interface ($\theta_{l-g}$ or $\theta_{g-l}$) and  liquid-liquid interface ($\theta_{l-l}$). In stratified flow model, the interface flux between different phases can be calculated at the cell boundaries. The AUSM$^+$-up scheme was introduced for the liquid-liquid or gas-gas flux calculation at cell interface and the exact Riemann solver was used to calculate the liquid-gas flux at cell interface. More details of the method can be found in \cite{Chang2007}.

\subsection{Configuration setup}

In this section, we introduce the configuration of calculation domain. It is a two-dimensional (2D) rectangle ($l_1 \times l_2$). Usually the length $l_1$ should cover several wavelength, and the width $l_2$ is equal to several bubble diameter. The calculation domain was designed with the two-phase mixture zone in the center and the single-phase zone at both sides. %, as illustrated in Fig. \ref{fig03}.

\begin{figure}
\vspace*{0.0cm}
\hspace*{2.3cm}
\includegraphics[scale=0.5]{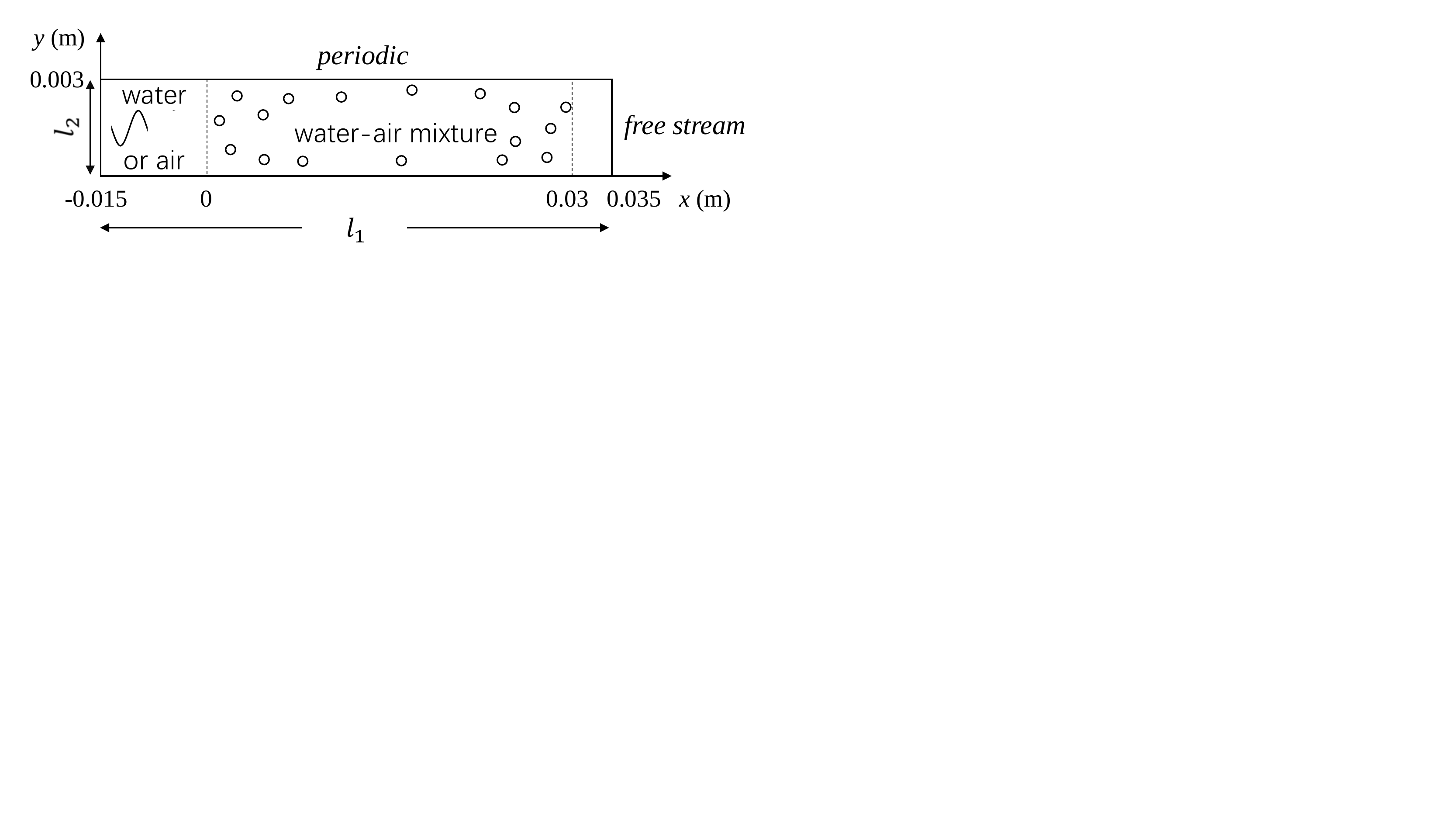}
\vspace*{-7.0cm}
\caption{2D calculation configuration}
\label{fig03}
\end{figure}

Figure \ref{fig03} is a typical configuration in our simulation case. The liquid-gas mixture zone locates in $30\times 3$ mm$^2$. Both top and bottom edges were applied for periodic boundary condition. The inlet was setup with the perturbation condition as we introduced in Eqn. \ref{perturbationBC}, and outlet with free stream boundary condition. More specifically, in case of liquid continuum, the inlet condition is applied with perturbation condition of water. And in case of gas continuum, the perturbation condition of air was applied for inlet boundary condition. The dispersed phase, either liquid or gas, is randomly distributed in the mixture zone with the same diameter. The distance between centers of two arbitrary bubbles/droplets should satisfy the following relationship to ensure well-distributed particles,
\begin{equation}
l \ge \epsilon D
\end{equation}
where $\epsilon > 1.0$ is a controlling parameter, for example $\epsilon = 1.03$ if $\alpha_g=0.5$ and $\epsilon = 1.3$ if $\alpha_g=0.3$. In the other way, the particles are regularly aligned in the mixture zone for comparison. The number of particles is calculated by
\begin{equation}
N=\frac{4\alpha_d S}{\pi D^2}
\end{equation}
where $S$ is the area of mixture zone.

The DIM is used in our work. The interface separating the two phases is captured by $\phi$, which is defined as a signed distance from the interface. The negative sign is chosen for the dispersed phase and the positive sign for the continuous phase,
\begin{equation}
{\alpha_d}=
\begin{cases}
 0 & \text{if}\ {\phi}\geq 0.5 h \\
 1 & \text{if}\ {\phi}\leq -0.5 h \\
 \displaystyle \cos \left[\frac{\pi}{4} \left(1+\frac{2\phi}{h}\right)\right] & \text{if}\ |{\phi}|< 0.5h
\end{cases}
\end{equation}
Here $h$ is the thickness of interface. In our work, $h$ is chosen for two-grid size,
\begin{equation}
h=2\Delta x
\end{equation}

It is very expensive for simulation at low frequency $f=1$ kHz. In this case, there are 1.24 million grids in a typical 2D configuration. And it takes 2.6 million time steps for a simulated time 1.75 ms, which costs about 66600 core-hours. Due to the limitation of our calculation resources, part of our work is done at $f=10$ kHz, although \cite{Karplus1958} did the reference experiment at lower frequency $f \sim 1$ kHz. It saves calculation time at higher frequency (but still below natural frequency of bubble $\omega_n$) in the following two aspects: 1) the wavelength of higher frequency is shorter and thus we may design a smaller mixture zone to cover typical length scale of several wavelength. 2) the simulated physical time which covers several period of wave could also be reduced. Thus, the number of time steps decrease considering the Courant-Friedrichs-Lewy (CFL) condition.

It was not easy to find the original article about the reference experiment at first. We could only find the limited information from \cite{Brennen2005}. Therefore, part of simulation was done at $D=0.15$ mm, which we considered as the most probable value from our experiences. Fortunately, the value is very close to the reported value $D=0.1$ mm by \cite{Karplus1958}.

\section{Results and discussion}\label{results}
In this section, we first present the appropriate simulation configuration, such as domain and meshes. The convergence of methods is studied. Second, the momentum analysis is given during wave propagation in two-phase flow. The homogeneous flow condition is discussed in this part. Third, the bubble thermodynamics are studied. The bubble behavior (isothermal or adiabatic) are discussed in low and high frequency (still below $\omega_n$). Finally, the acoustic dispersion is studied.

\subsection{Convergence of numerical methods and sensitivity study}
%[宽度的影响] fig061, fig3
%[网格收敛性] fig062, fig04(正在补算一个)
%[dB的影响]  fig062
%[气泡分布的影响] fig05

%[宽度的影响] fig061, fig3
%As we mentioned before, in order to save the time, we tried to reduce the calculation domain either by its width. On the other hand, the width $l_2$ should be large enough to eliminating the unanticipated effects from the periodic boundary condition. The sensitivity study of width $l_2$ is shown in Fig. \ref{fig3}. The result shows that it is reasonable to set
%\begin{equation}
%l_2 = 5D
%\end{equation}
%in our calculation.

\begin{figure}
\vspace*{-5.0cm}
\hspace*{1.0cm}
\includegraphics[scale=0.5]{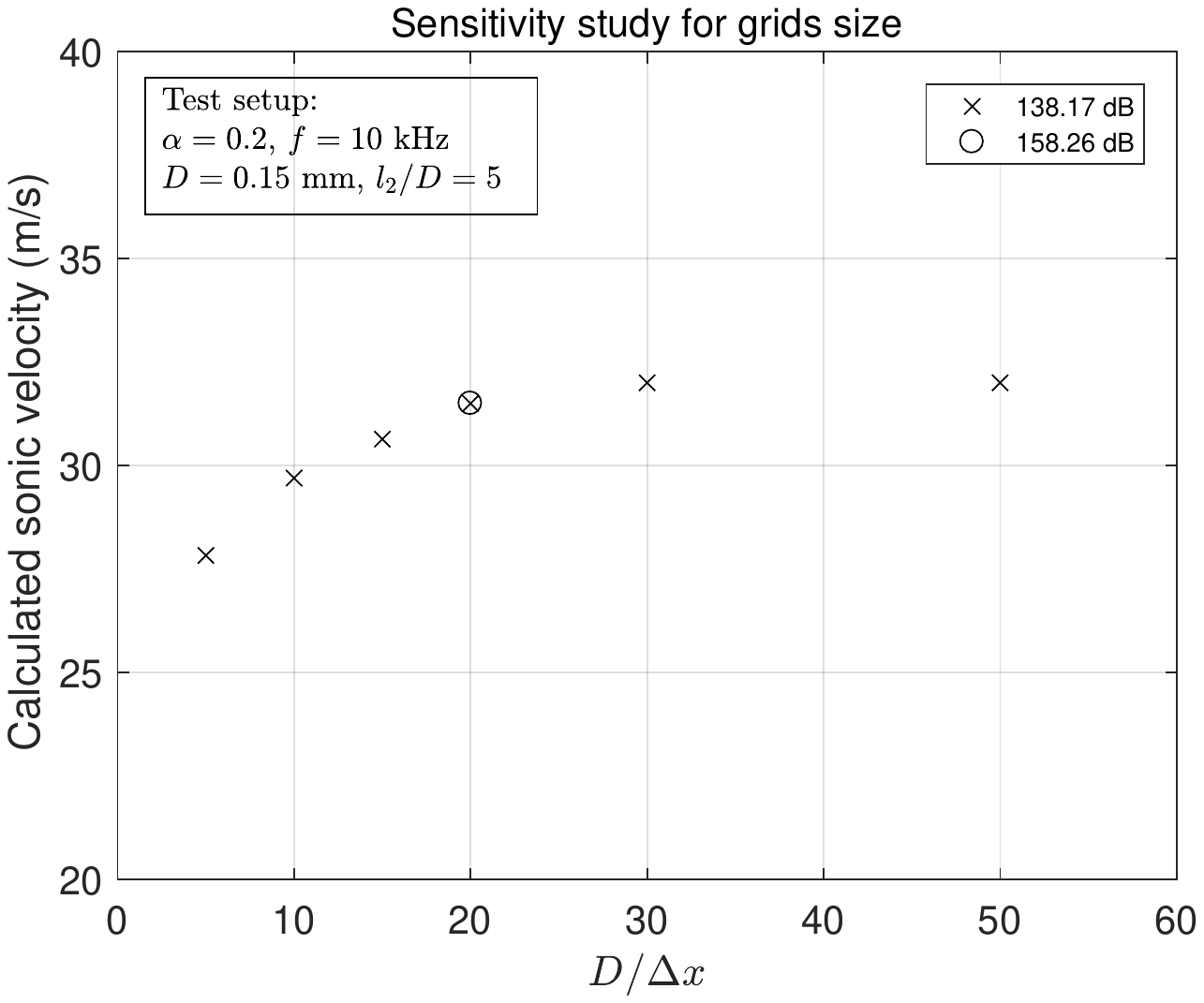}
\vspace*{-5.0cm}
\caption{Sensitivity test for grid resolution and sound intensity $L_p$}
\label{fig04}
\end{figure}

%[网格收敛性] fig062, fig04(正在补算一个)
First, considering the calculation efficiency, appropriate size of domain and grid should be determined. Considering the periodic boundary condition in $y$ direction, $l_2=5D$ is used in this work. The reasonable grid size $\Delta x$, which should be sufficient to capture the gas-liquid interface, is chosen to save the calculation time. The sensitivity study about the resolution $D/\Delta x$ is shown in Fig. \ref{fig04}. The result shows that the wave propagation speed has the first order convergence. Thus, it is reasonable for us to choose $D/\Delta x =20$ in the calculation.
%[dB的影响]  fig062
We also did the sensitivity study to the intensity of sound $L_p$. It is found that sometimes the wave was too weak to be captured. Thus, we used in our simulation the strong shock wave condition as listed in Table. \ref{postshock}. Obviously, we obtain the same speed of sound for both weak and strong condition, as shown in Fig. \ref{fig04}.

\begin{figure}
\vspace*{-0.5cm}
\hspace*{-0.2cm}
\includegraphics[scale=0.45]{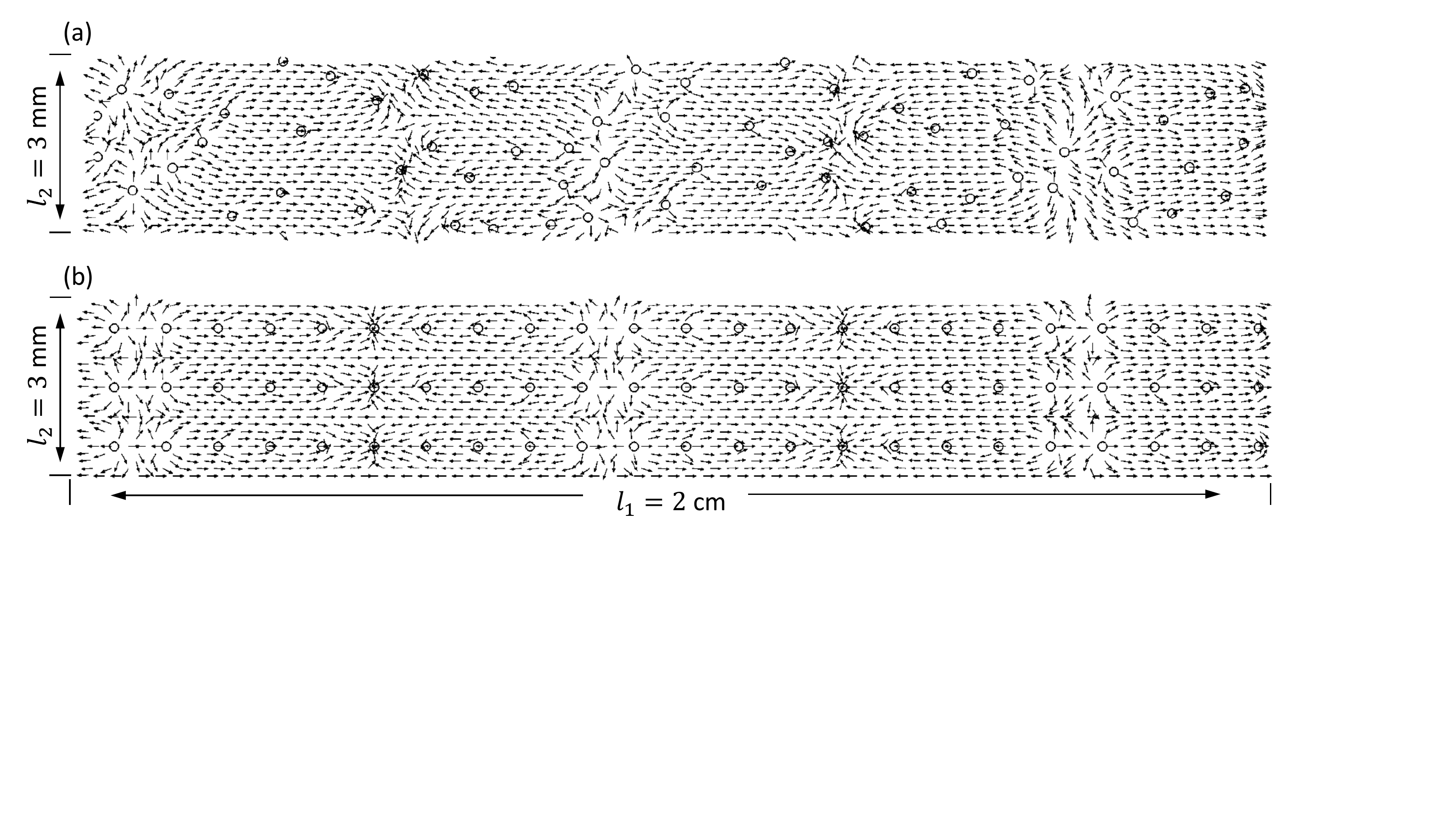}
\vspace*{-3.5cm}
\caption{Comparison of velocity field between (a) randomly distributed and (b) aligned gas bubbles}
\label{fig05}
\end{figure}

%[气泡分布的影响] fig05
The distribution pattern of particles is studied in this work. Figure \ref{fig05} shows the velocity field of flow in different distribution pattern of dispersed phase, at the same time $t=3\times 10^{-4}$ s, where $\alpha=0.02$, $D=0.15$ mm and $f=10$ kHz. The similar two regular flow patterns indicate the waves almost have the same propagation speed. Further investigation shows that $c=76.4$ m/s for random pattern in Fig. \ref{fig05}a and $c=75.3$ m/s for aligned pattern in Fig. \ref{fig05}b.

In summary, we choose $l_2=5D$, $D/\Delta x=20$, $L_p=158.26$ dB (water) and $L_p=158.48$ dB (air) for the simulation. All the dispersed particles are randomly distributed in calculation domain.

\subsection{Sonic speed analysis with momentum}\label{c_momentum}
\begin{figure}
\vspace*{-4.0cm}
\hspace*{1.0cm}
\includegraphics[scale=0.5]{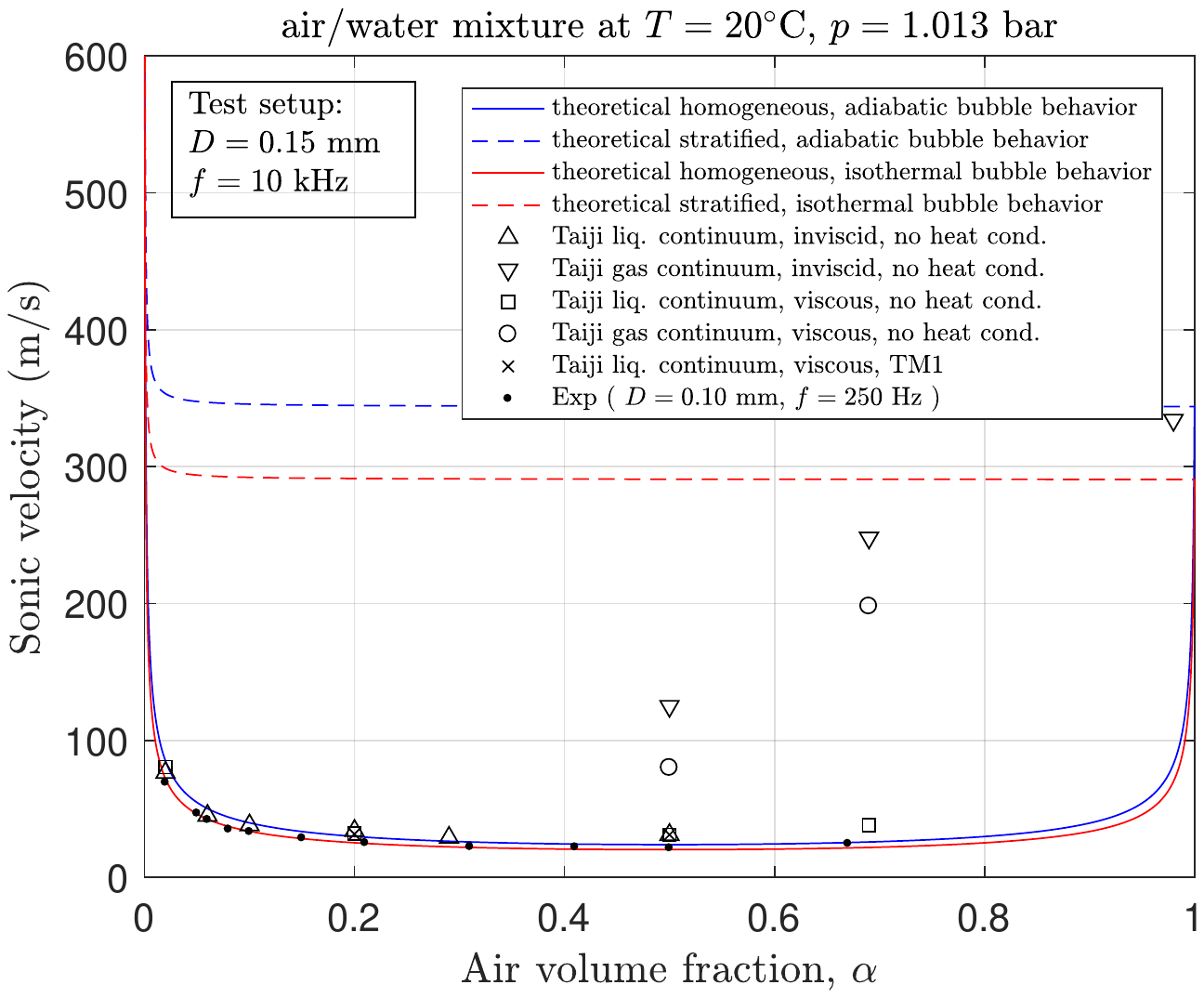}
\vspace*{-5.0cm}
\caption{The speed of sound in a bubbly air/water mixture at atmospheric pressure}
\label{fig06}
\end{figure}

We calculated the speed of sound in case of $f=10$ kHz and $D=0.15$ mm, as shown in Fig. \ref{fig06}. Note TM1 refers to Eqn. \ref{ek}. It is found that when $\alpha \rightarrow 0$, $c \rightarrow c_{\rm hom}$. Otherwise when $\alpha \rightarrow 1$, $c \rightarrow c_{\rm sep}$.

Consider a general steady fluid flow characterized by a velocity, $U$, and a typical dimension, $l$. A particle in this flow will experience a typical fluid acceleration of $U^2/l$ for a typical time $l/U$ and hence will develop a velocity, $W$, relative to the fluid. The maximum value of $W$, denoted by $W_m$, is recognized as terminal velocity.
%For a given circumstance, one must first consider whether the available time, $l/U$, is larger or smaller compared with the typical time, $t_u$, required for the particle to reach its terminal velocity.
For a given circumstance, one must first compare the available time ($l/U$), with the typical time required for the particle to reach its terminal velocity ($t_u$).
If $t_u \ll l/U$, we refer this as the quasistatic regime. On the other hand, if $t_u \gg l/U$, we call it as transient regime.
\cite{Brennen2005} proposed two non-dimensional numbers (p.78),
\begin{equation}\label{X}
X=\frac{D}{2l} \left|1-\frac{\rho_d}{\rho_c}\right|
\end{equation}
and,
\begin{equation}\label{Y}
Y= \left.{\left|1-\frac{\rho_d}{\rho_c}\right|}   \middle/ {\left(1+\frac{2\rho_d}{\rho_c}\right)}\right.
\end{equation}
%where $l$ is a typical dimension, $\rho_d$ is the density of dispersed phase and $\rho_c$ is the density of continuous phase.
For inviscid dispersed flow, a quasistatic regime is suggested,
\begin{equation}\label{homocondition}
X \ll Y^2
\end{equation}
In this regime, the relative motion between phases could be neglected and the flow is in homogeneous condition.

In dispersed air/water two-phase flow, the typical length is wavelength. From Eqns. \ref{X}, \ref{Y} and \ref{homocondition}, the homogeneous condition is satisfied if
\begin{equation}\label{ourhomocondition}
\frac{D}{2\lambda} \left(1-\frac{\rho_g}{\rho_l}\right) \ll 1
\end{equation}
Take a typical simulation case for the dispersed air/water mixture, $D = 1.0\times10^{-4}$ m, and $\lambda = 2.5 \times 10^{-3} $ m. In this condition, Eqn. \ref{ourhomocondition} is satisfied and the flow is homogeneous.

However, in dispersed water/air two-phase flow, the density of water droplet is much larger than that of air. The homogeneous condition is not satisfied any more. From Fig. \ref{fig06}, we have $c_{\rm hom}<c<c_{\rm sep}$ (here $c$ is the simulated speed of sound in gas continuum), which indicates that the relative motion could not be ignored. One interesting thing could be found in the case $\alpha =0.5$. With the same volume fraction, the simulated speed of sound in liquid continuum is different from the one in gas continuum. It is due to the shape effects on drag. In dispersed air/water mixture, the gas bubbles are spherical. They are easily accelerated by drag force due to their light density. Thus, the terminal velocity could be reached instantly and we say that the two-phase flow is in the quasistatic regime. However, in the dispersed water/air mixture, the liquid droplets are spherical. The acceleration of the droplet is not so easy and it takes a long time for droplet to reach the terminal velocity. The two-phase flow is then in the transition regime and homogenous condition is not satisfied.

\subsection{Sonic speed analysis with bubble thermodynamics}
\label{speed-thermodynamics}
Due to large heat capacity, $T_l$ is almost constant at the equilibrium temperature $T_e=293.15$ K,
\begin{equation}
T_l \approx T_e
\end{equation}
In TM1, $T_g$ could be varied from $T_e$ due to the compression and rarefaction accompanying the passage of the sound wave. In what follows, we discuss the heat diffusion problem which is important to bubble thermodynamics in two-phase flow.

\begin{figure}
\vspace*{-0.5cm}
\hspace*{3.0cm}
\includegraphics[scale=0.5]{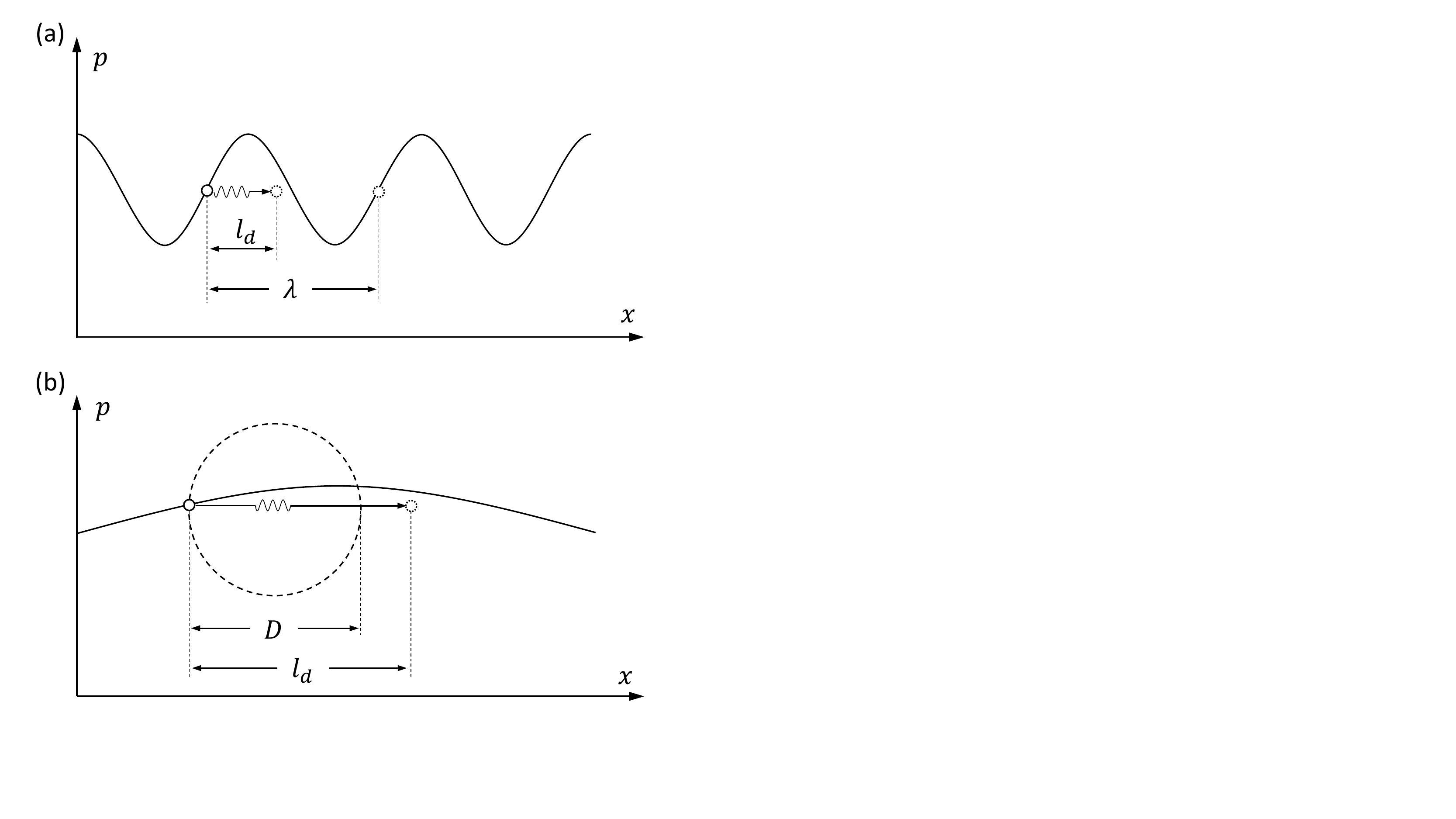}
\vspace*{-1.5cm}
\caption{Illustration of gas thermodynamics in wave propagation: diffusion of thermal energy in (a) single-phase flow and (b) two-phase flow}
\label{fig07}
\end{figure}

First, we consider a 1D heat conduction problem in single-phase flow, and heat diffusion in two-phase flow is quite analogous.
\begin{equation}
\frac{1}{r^2} \frac{\partial}{\partial r}(r^2 \frac{\partial T}{\partial r})=\frac{1}{\kappa} \frac{\partial T}{\partial t}
\end{equation}
The analytic solution can be found in \cite{Ernesto2006}. The characteristic diffusive length
\begin{equation}
l_d = \sqrt{\kappa \tau}
\end{equation}
is introduced during the typical time $\tau=1/f$. Comparing $l_d$ with the wavelength $\lambda=c/f$, as shown in Fig. \ref{fig07}a, the relaxation frequency
\begin{equation}
f_{\rm tc,1\Phi}= \frac{c^2}{ \kappa}
\end{equation}
for thermal conduction is introduced to determine fluid condition during wave propagation: isentropic or isothermal.
\cite{Dijk2005} pointed out that for single-phase flow in the low frequency $f\ll f_{\rm tc,1\Phi}$, the equilibrium speed of sound is given as
\begin{equation}
c_s^2=\left(\frac{{\rm d}p}{{\rm d}\rho}\right)_s
\end{equation}
for isentropic condition. In the high frequency $f\gg f_{\rm tc,1\Phi}$, the conduction of heat fully dominates the energy balance. In this case the frozen speed of sound is given by,
\begin{equation}
c_T^2=\left(\frac{{\rm d}p}{{\rm d}\rho}\right)_T
\end{equation}
for isothermal condition.

The above discussion is valid for single-phase flow. For dispersed two-phase flow, considering bulk modulus of liquid is usually much larger than that of gas, the acoustics are majorly dependent on the gas thermodynamics. In air/water two-phase flow, the accumulated gas temperature due to compression or decompression is diffused into the surrounding liquid from the interface, as shown in Fig. \ref{fig07}b. The critical frequency is then given by
\begin{equation}\label{ftctp}
f_{\rm tc,TP}=\frac{\kappa_d}{ D^2}
\end{equation}
where $\kappa_d$ is the thermal diffusivity of dispersed phase. For air/water mixture at $T_e=293.15$ K, $\kappa_{g}=1.9\times10^{-5}$ m$^2$/s, $D = 1\times 10^{-4}$ m, thus $f_{\rm tc,TP} \sim 1$ kHz. The air is in the isothermal condition during wave propagation if $f\ll 1$ kHz, which is different from single-phase flow. In section \ref{c_dhf}, we mentioned that $\gamma_g =1.0$ for isothermal condition and $\gamma_g =1.4$ for adiabatic condition. Here $f_{\rm tc,TP}$ indicates which condition, isothermal or adiabatic, that bubbles experience. In air/water mixture, we have
\begin{equation}
{\gamma_g}=
\begin{cases}
 1.0 & \text{if}\ f\ll f_{\rm tc,TP} \\
 1.4 & \text{if}\ f\gg f_{\rm tc,TP} \\
\end{cases}
\end{equation}

\begin{figure}
\vspace*{-5.0cm}
\hspace*{1.0cm}
\includegraphics[scale=0.5]{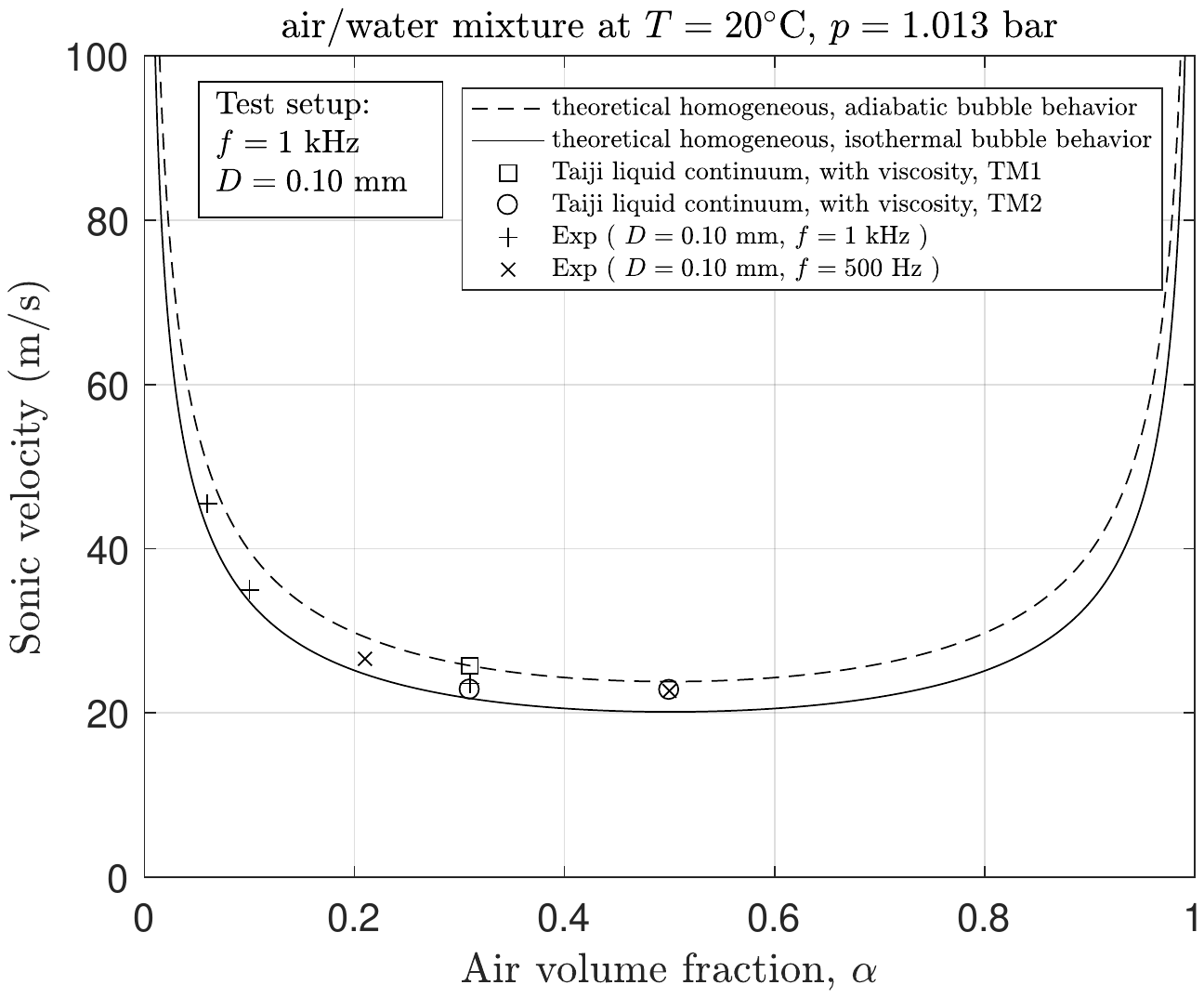}
\vspace*{-4.5cm}
\caption{The speed of sound in a bubbly air/water mixture at atmospheric pressure}
\label{fig08}
\end{figure}

Figure \ref{fig08} shows the speed of sound in air/water mixture with $f=1$ kHz. TM1 and TM2 refer to Eqns. \ref{ek} and \ref{ek3} respectively. First, the result shows that with $f=1$ kHz, the calculated sonic velocity agrees better with the theoretical prediction, compared with $f = 10$ kHz. It could be explained by the homogeneous condition in low frequency, as indicated in Eqn. \ref{ourhomocondition}. Second, consider the case with $\alpha=0.31$. According to the experimental measurement, the simulation result in TM2 is better than in TM1. It shows that the gas bubbles are more favorable in isothermal condition than in adiabatic condition. We introduced the heat transfer between two phases in TM1. However further investigation of the gas temperature shows that the bubbles have a small temperature variation either by compression or decompression. Therefore, we consider TM1 simulate an adiabatic process roughly. The result with $\alpha=0.5$ also shows an isothermal bubble behavior during wave propagation.

%色散关系
\subsection{Acoustic dispersion}
The acoustic dispersion relation was derived from the linearized conservation equations and the Rayleigh equation during the past few decades (\cite{Mecredy1972,Ardron1978,Cheng1983,Cheng1985,Ruggles1988,Ruggles1989,Drui2016}). In these work, a 1D two-fluid model was used to predict the measurement. The model includes bubble dynamics, viscous flow effects and interfacial heat transfer. The dependent variables in this model are space/time averaged variables. As \cite{Cheng1985} discussed, the length scale is large compared to bubble radius and the inter-bubble distance but is small compared to the wavelength, which makes difference from DNS method.
In quiescent two-phase flow, the pressure drag force could be ignored since the relative velocity is zero. The virtual mass force dominates the interfacial momentum transfer.  \cite{Cheng1983} studied the effect of the virtual mass coefficient $c_{\rm VM}$. In his work, the frequency dependent phase velocity is given by,
\begin{equation}\label{DRCheng}
c_{\rm ph} = \left[\frac{\left(\displaystyle \frac{\alpha_g}{\rho_g c_g^2 (1-\omega^2/\omega_{\rm n}^2)}+ \frac{\alpha_l}{\rho_l c_l^2} \right)
\left(\rho_g + \bar{\rho}\displaystyle \frac{c_{\rm VM}}{\alpha_l}\right)}
{\displaystyle \frac{\alpha_l \rho_g}{\rho_l} + \frac{\alpha_g}{1-\omega^2/\omega_{\rm n}^2} + \left(1+\frac{\alpha_g}{\alpha_l(1-\omega^2/\omega_{\rm n}^2)}\right)c_{\rm VM}} \right]^{-1/2}
\end{equation}
where
\begin{equation}
\bar{\rho}=\alpha_g \rho_g + \alpha_l \rho_l
\end{equation}
is the average density and
\begin{equation}
\omega_{\rm n} = \frac{2}{D} \left( \frac{3\gamma_g p}{\rho_l} + \frac{12\gamma_g \sigma -4\sigma}{\rho_l D}  \right)^{1/2}
\end{equation}
is the natural frequency of a pulsating bubble. The evaluation of Eqn. \ref{DRCheng} leads to the typical sonic velocity,
\begin{equation}
{c_{\rm ph}}=
\begin{cases}
 c_l & \text{if}\ \omega \rightarrow \infty, \alpha_g \ll 1 \\
 c_{\rm hom} & \text{if}\ \omega \ll \omega_n, c_{\rm VM} \rightarrow \infty\\
 c_{\rm sep} & \text{if}\ \omega \ll \omega_n, c_{\rm VM} =0\\
\end{cases}
\end{equation}

\cite{Drui2016} proposed a two-fluid model that accounts for two-scale kinematic effects: bulk kinematics and small-scale vibrations. Two relaxation parameters related to mechanical equilibrium between materials are identified in the model: micro-inertial $\nu$ and micro-viscosity $\varepsilon$. The evaluation of $\nu$ and $\varepsilon$ is notable as it could be replaced by infinitely fast relaxation processes as studied in \cite{Saurel2009,Shyue2014}. The 4-equation model is derived for $\nu \rightarrow 0$ and $\varepsilon = O(1)$, in which the dispersion relation is given as,
\begin{equation}\label{DRDrui}
\left(\frac{k^\varepsilon(\omega)}{\omega}\right)^2 = \frac{i\varepsilon \omega+ c^{-2}_{\rm hom} H }{i\varepsilon c^2_{\rm Frozen} \omega +H}, \:\:\:\: c_{\rm ph}^\varepsilon(\omega)=\operatorname{\mathbb{R}e} \left[ \frac{\omega}{k^\varepsilon(\omega)}\right]
\end{equation}
where
\begin{equation}
c^2_{\rm Frozen} = \frac{\alpha_g \rho_g}{\bar{\rho}}c_g^2 +\frac{\alpha_l \rho_l}{\bar{\rho}}c_l^2
\end{equation}
and
\begin{equation}
H = \frac{ \rho_g \rho_l c_g^2 c_l^2}{\alpha_g \alpha_l \bar{\rho}}
\end{equation}

The acoustic dispersion is also studied in our work. It can be related to the following two aspects. The first is the non-equilibrium effects. \cite{Brennen2005} discussed that when the acoustic excitation (or driving) frequency approaches the natural (or resonant) frequency of the bubbles, the bubbles are not in dynamic equilibrium. The EOS we used to calculate $c$ in single-phase fluid (such as Eqn. \ref{c_iso}) does not establish any more. \cite{Karplus1958} estimated that a bubble of 0.1 mm in diameter has the natural frequency $f=55$ kHz. \cite{Fox1955} calculated the sound velocity in air/water mixture in function of frequency. In their conclusion, it shows that very little dispersion is expected in our case since the operation frequency ($\sim1$ kHz) is far below the resonant frequency ($\sim55$ kHz).

\begin{figure}
\vspace*{-5.0cm}
\hspace*{1.0cm}
\includegraphics[scale=0.5]{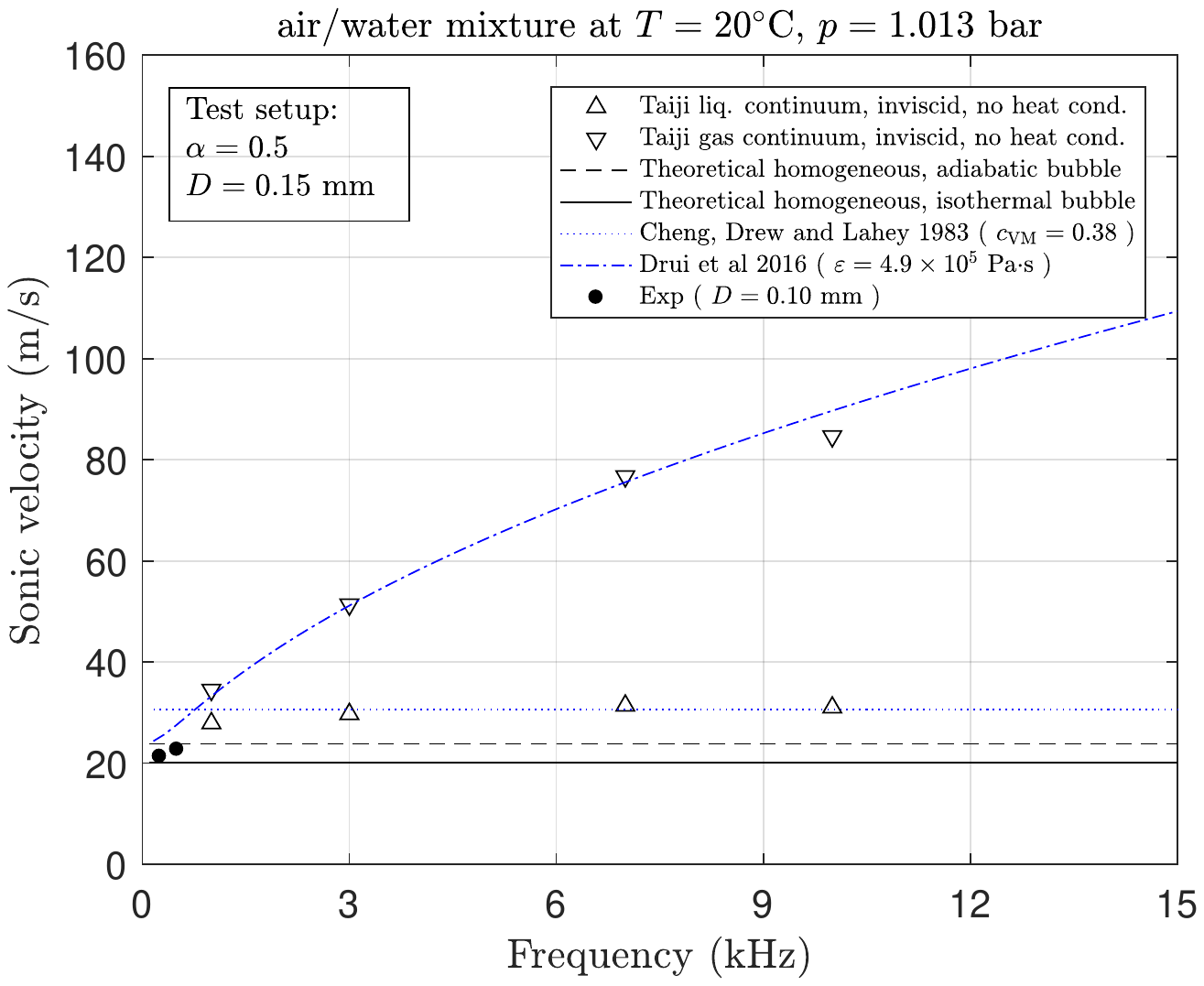}
\vspace*{-4.5cm}
\caption{The acoustic dispersion in air/water mixture}
\label{fig09}
\end{figure}

The second reason comes from the relative motion which is ignored in the homogeneous flow model. That is the explanation for dispersion below natural frequency $\omega_n$. In fact, Eqn. \ref{ourhomocondition} is not satisfied with high frequency. \cite{Wijngaarden1976} formulated the sound velocity in an approximate manner when there exists relative motion. In his formulation, there will be a larger sound velocity if relative motion exists. Our simulation results in Fig. \ref{fig09} agree with the conclusion. In our method, Eqn. \ref{moment} promises there could be sufficient momentum exchange between phases, but the equality of phasic velocity ${\mathbf u}_k$ is not necessary.

In the scope of low frequency ($\omega \ll \omega_n$), our simulation result can be compared with the analytical results as introduced in Eqns. \ref{DRCheng} and \ref{DRDrui}. The simulation results and analytical results agree well with each other, as shown in Fig .\ref{fig09}. In gas dispersed two-phase flow, we applied the \cite{Cheng1983} model and chose $c_{\rm VM}=0.38$ in Eqn. \ref{DRCheng}. While in droplet dispersed two-phase flow, the \cite{Drui2016} model was applied and $\varepsilon=4.5\times10^5$ Pa$\cdot$s is set in Eqn. \ref{DRDrui}.

\section{Conclusion}
In this work, we provided an efficient direct numerical simulation method to study the speed of sound in air/water two-phase flow. The diffuse-interface method is used to capture liquid-gas interface. The stratified flow method is adopted for flux calculation in volume of fluid method.
The grid convergence was studied and the result indicates first order convergence.
The distribution pattern of particles was also studied and it shows the speed of sound is irrelevant to the pattern as long as the particles are well distributed. The simulation results were compared with both theoretical and experimental results.
The study shows that the speed of sound in two-phase flow is dependent on both momentum relaxation and thermal energy relaxation between the two phases. More specifically, first, the speed of sound in two-phase flow relies on the frequency.
In low frequency, the two-phase flow is in homogeneous condition.
The speed of sound greatly decreases in large volume fraction. As frequency increases, the relative motion exists and thus speed of sound increases. Second, the shape effects on drag also influence on speed of sound. For example, the speed of sound varies largely in different continuous phase. Third, the study of bubble thermodynamics shows that unlike in single-phase flow, the air is in isothermal condition in air/water two-phase flow in low-frequency waves. Finally the simulation dispersion relation agrees with the analytical results in the low frequency regime.

%an isothermal bubble behavior in air/water mixture during the wave propagation

%the shape effects on drag h

% is studied using the stratified flow method. The two-dimensional domain was used for DNS calculation.

% at both high frequency $f=10$ kHz and low frequency $f=1$ kHz.

%The sensitivity study on domain size and resolution shows that our simulation results are grid independent.

%The calculation result shows that the speed of sound in two-phase flow relies on the frequency. Typically, we have large speed of sound at high frequency in two-phase flow. The homogeneous flow condition for phasic velocity is satisfied at low frequency in the study. Meanwhile the simulation result indicates an isothermal bubble behavior in air/water mixture during the wave propagation.

\section{Acknowledgments}
This work was supported by the National Natural Science Foundation of China (Grant Nos. 91230203, 11202020, U1530401), the President Foundation of Chinese Academy of Engineering Physics (Grant No. 201501043) and the China Postdoctoral Science Foundation (Grant No. 2016M591059). We acknowledge the computational supports from the Special Program for Applied Research on Super Computation of the NSFC-Guangdong Joint Fund (the second phase) under Grant No.U1501501, and from the Beijing Computational Science Research Center (CSRC). The authors also thank Dr. Chih-Hao Chang for the fruitful supports and helpful discussions.
\cleardoublepage

\begin{supertabular}{ll}

\multicolumn{2}{l}{\textbf{Nomenclature}}\\
&\\
$A$ & area of cross section, m$^2$ \\
$c$ & speed of sound, m$\cdot$s$^{-1}$ \\
$c_{p}$ &       specific heat capacity, J$\cdot$kg$^{-1}\cdot$K$^{-1}$\\
$D$&     diameter, m \\
%Eo      &Eotvos number \\
$e$& internal energy, J$\cdot$kg$^{-1}$ \\
%$F$&force\\
$f$     &   driving frequency, Hz \\
$h$& enthalpy, J$\cdot$kg$^{-1}$; or interface thickness, m \\
$K$&bulk modulus, N$\cdot$m$^{-2}$ \\
$k$ & thermal conductivity, W$\cdot$m$^{-1}\cdot$K$^{-1}$ \\
$L_p$&sound pressure level, dB\\
$l$&length, m\\
$M$& Mach number \\
$p$ & pressure, Pa\\
$p_0$ & reference pressure, Pa\\
$Q$ & heat, J; or thermodynamic constraint\\
$ {\mathbf q}^{\prime\prime}$&heat flux, W$\cdot$m$^{-2}$ \\
$R$ & specific gas constant, J$\cdot$kg$^{-1}\cdot$K$^{-1}$\\
$S$& area, m$^2$ \\
$T$&temperature, K\\
%$T_0$& equilibrium temperature, K\\
$t$       &  time, s\\
$t_u$       &  typical time for particle to reach its terminal velocity, s\\
$U$&typical velocity, m$\cdot$s$^{-1}$\\
${\mathbf u}, u$&velocity, m$\cdot$s$^{-1}$\\
$W$& interface velocity or relative velocity, m$\cdot$s$^{-1}$\\
%$u_\tau$& friction velocity, m$\cdot$s$^{-1}$\\
%We& Weber number\\
$X$&non-dimensional number\\
$Y$&non-dimensional number\\
$\Delta x$ &grid size, m\\
&\\
\multicolumn{2}{l}{\emph{Greek letters}}\\
$\alpha$ & volume fraction\\
%$\beta$ & void fraction for continuous phase\\
%$\epsilon$&turbulent dissipation rate, m$^2\cdot$s$^{-3}$\\
$\epsilon$& parameter \\
%$\Gamma$&rate of phase change, kg$\cdot$m$^{-3}\cdot$s$^{-1}$\\
$\gamma$& specific heat ratio \\
$\kappa$    &           thermal diffusivity, m$^2\cdot$s$^{-1}$\\
$\lambda$ &   wavelength, m\\
$\mu$      & dynamic viscosity, kg$\cdot$m$^{-1}\cdot$s$^{-1}$\\
%$\nu$   &   kinematic viscosity, m$^{2}\cdot$s$^{-1}$ \\
$\nu$   & micro-inertia, kg $\cdot$m$^{-1}$ \\
$\omega$ & angular frequency, s$^{-1}$\\
$\phi$ & signed distance, m \\
%$\psi$  &  factor depending on bubble shape \\
$\rho$ &    density, kg$\cdot$m$^{-3}$\\
$\sigma$ &  interfacial tension, N$\cdot$m$^{-1}$ \\
$\boldsymbol \tau$     &  stress tensor, N$\cdot$m$^{-2}$\\
$ \tau$     &  typical time, s\\
%$\xi$& ratio\\
%$\tau_w$&wall shear stress, N$\cdot$m$^{-2}$\\
%$\theta$    &   contact angle\\
$\varepsilon$ & micro-viscosity, Pa$\cdot$s \\
&\\

\multicolumn{2}{l}{\emph{Superscripts}}\\
%eff & effective\\
%$d$ & dispersed or diffusive \\
%$l$ & lift\\
%$t$ & turbulence\\
%$td$ & turbulent dispersion\\
%$vm$ & virtual mass \\
%$w$ & wall \\
%$wl$ & wall lubrication\\
$\prime$ & perturbation \\

&\\

\multicolumn{2}{l}{\emph{Subscripts}}\\
1&dispersed phase or wave front\\
1$\Phi$& single-phase \\
2&continuous phase or post wave\\
%$1\Phi$&single phase\\
%BB& bubble breakup\\
%BC& bubble coalescence\\
%$b$& bubble\\
%$c$& convection\\
$c$& continuous \\
$d$ & dispersed or diffusive \\
%$fc$  & single phase forced convection \\
%$d$  & bubble departure \\
$e$  & equilibrium \\
%$ev$& evaporation\\
$g$& gas\\
hom& homogeneous\\
$i$& interphase\\
$k$ & phase\\
$l$& liquid\\
%$lo$& lift-off\\
m&maximum\\
n& natural \\
%$n$& normal or $n$th\\
%NUC& nucleation \\
ph&phase\\
%$p$&pressure\\
%$q$&quenching\\
%$r$&relative\\
%ref&reference\\
%sat&saturation\\
$s$&isentropic\\
sep&separated\\
$T$&isothermal \\
TP& two-phase \\
tc&thermal conduction\\
%$v$&vapor\\
%$w$& wall\\
$\infty$ & far field \\
\end{supertabular}




\appendix
\section{The derivation of equation in 1D compressible flow}
\label{1Dru}

\begin{figure}
\vspace*{-0.5cm}
\hspace*{2.5cm}
\includegraphics[scale=0.6]{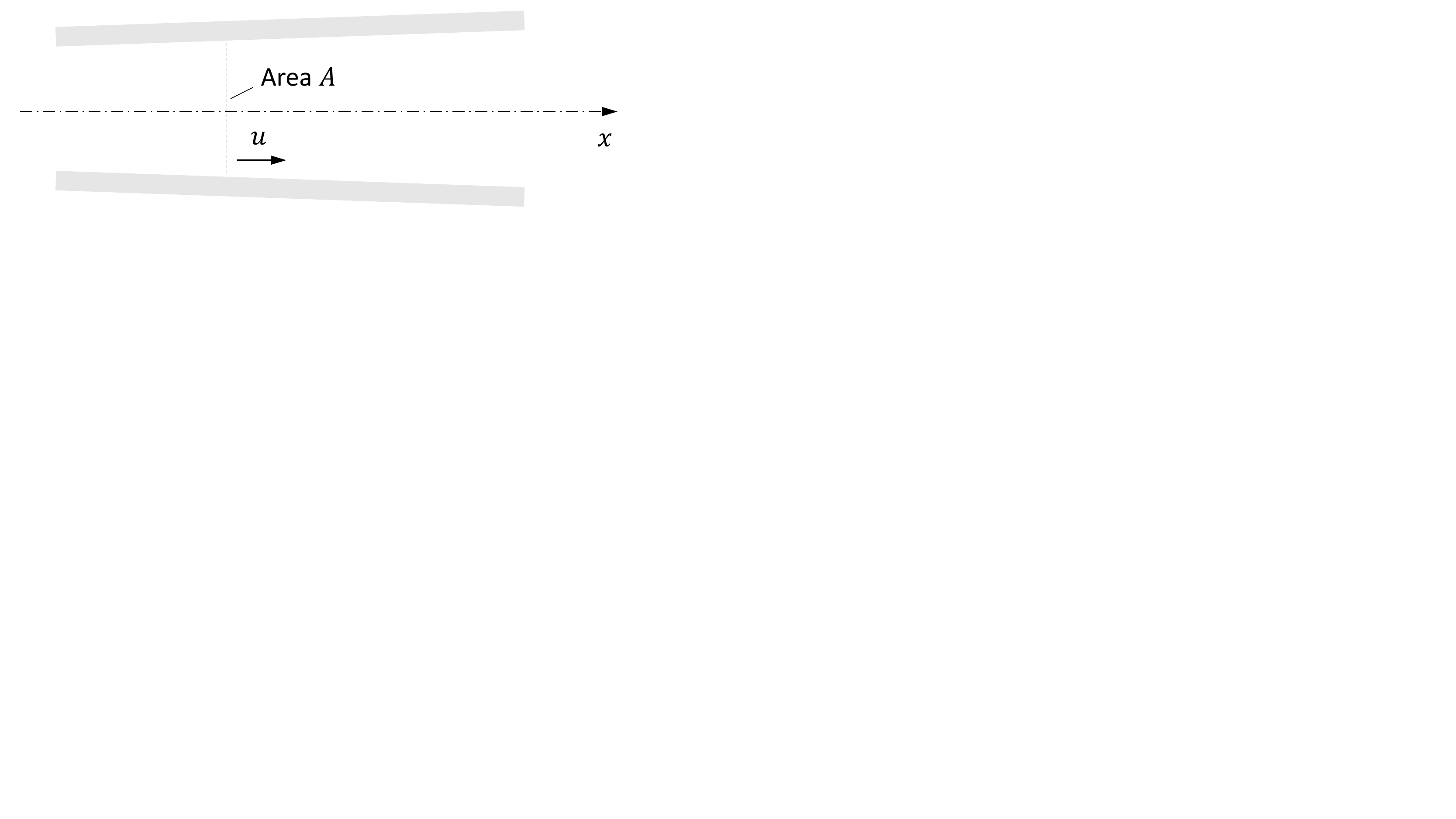}
\vspace*{-8.0cm}
\caption{One-dimensional flow}
\label{fig10}
\end{figure}

Consider a 1D inviscid flow as show in Fig. \ref{fig10}, the continuity equation can be written down as,
\begin{equation}\label{dru}
A \frac{\partial \rho}{\partial t} +\frac{\partial}{{\partial}x} \left( \rho Au  \right) =0
\end{equation}
where $u$ is the averaged flow velocity along $x$ axis and $A$ is the cross-section area.

The momentum equation can be written down as,
\begin{equation}\label{1DNS}
A \frac{\partial }{\partial t} \left(\rho u\right) + \frac{\partial}{\partial x}\left(\rho Au^2 \right) = - A\frac{\partial p}{\partial x}
\end{equation}
Combining Eqns. \ref{dru} and \ref{1DNS}, we simply obtain
\begin{equation}\label{1DNS2}
\rho u  \frac{\partial u}{\partial x} = - \frac{\partial p}{\partial x}
\end{equation}
for steady flow. The detailed derivation of Eqns. \ref{dru} and \ref{1DNS} can be found in \cite{Hdaneshyar1976}.

Consider isentropic flow, we have the relationship
\begin{equation}
\frac{{\rm d}p}{p} = \gamma  \frac{{\rm d}\rho}{\rho}
\end{equation}
where $\gamma$ is the specific heat ratio. Using the ideal equation of state, we have
\begin{equation}
p = \rho  R  T
\end{equation}
where $R$ is the specific gas constant and $T$ is the absolute temperature. Thus we have,
\begin{equation}\label{dpdr}
{\rm d}p = c^2  {\rm d}\rho
\end{equation}
where the speed of sound,
\begin{equation}
c = (\gamma  R  T)^{1/2}
\end{equation}

Combining Eqns. \ref{1DNS2} and \ref{dpdr}, we have
\begin{equation}
-M^2 \frac {{\rm d} u}{u} = \frac{{\rm d}\rho}{\rho}
\end{equation}
where the Mach number
\begin{equation}
M = u/c
\end{equation}

\section{The derivation of speed of sound in two-phase flow}
\label{dcmix}
In this section, we derived the speed of sound in two-phase flow as in Eqns. \ref{chom} and \ref{csep}.

\subsection{Speed of sound in dispersed homogeneous flow}

Adding perturbations into Eqn. \ref{alpha1}, we have
\begin{equation}\label{alpha2}
 \frac{\partial \alpha_k^{\prime} } {\partial t}+ \frac{\alpha_k}{\rho_k} \frac{\partial \rho_k^{\prime} } {\partial t} =-\nabla \cdot (\alpha_k  \mathbf u_k^{\prime})
\end{equation}
if the convection term
\begin{equation}
\mathbf u_k^{\prime} \cdot \nabla \rho_k^{\prime}=0
\end{equation}
is ignored. Here we denote $\phi^{\prime}$ as the deviation term from the equilibrium $\phi$. We designate
\begin{equation}\label{EOSp1}
c_k^2 = \left( \frac{dp_k}{d\rho_k} \right)_Q
\end{equation}
where $Q$ is the thermodynamic constraint. As \cite{Brennen2005} suggested (p223), in most practical circumstances, we have the equilibrium local pressure for both components if the surface tension is neglected,
\begin{equation}\label{pk}
 p = p_1 = p_2
\end{equation}
%and
%\begin{equation}\label{pkp}
%p^{\prime}=p_1^{\prime}=p_2^{\prime}
%\end{equation}
Therefore, the subscript of pressure could be omitted.

Substituting Eqn. \ref{EOSp1} into Eqn. \ref{alpha2}, we have
\begin{equation}\label{alpha3}
 \frac{\partial \alpha_k^{\prime} } {\partial t}+ \frac{\alpha_k}{\rho_k c_k^2} \frac{\partial p^{\prime} } {\partial t} =-\nabla \cdot (\alpha_k  \mathbf u_k^{\prime})
\end{equation}
Then combining Eqn. \ref{alpha3} for both phases, we have
\begin{equation}\label{alpha4}
 \frac{\partial p^{\prime} } {\partial t}=-K \nabla \cdot  \mathbf u^{\prime}
\end{equation}
where we define the effective modulus of the two-phase medium $K$ as
\begin{equation}\label{K1}
 \frac{1 } {K}= \frac{\alpha_1 } {K_1}+  \frac{\alpha_2 } {K_2}
\end{equation}
Here $K_k=\rho_k c_k^2$ refers to the bulk modulus of each phase. And the bulk velocity fluctuation

\begin{equation}\label{u1}
\mathbf u^{\prime}= \alpha_1\mathbf u_1^{\prime}+ \alpha_2 \mathbf u_2^{\prime}
\end{equation}
The LHS of Eqn. \ref{alpha4} refers to the change of pressure, and RHS is related to the change of volume.

In homogenous flow, the interfacial forces $\mathbf F_{i}$ in Eqn. \ref{NS1} are so large that the relative velocity is neglected, thus we have
\begin{equation}
\mathbf u = \mathbf u_1 = \mathbf u_2
\end{equation}

Combining Eqn. \ref{NS1} for both phases, we have
\begin{equation}\label{NS2}
 \frac{\partial  \rho \mathbf u} {\partial t}+ \nabla \cdot ( \rho \mathbf u \mathbf u)=- \nabla p
\end{equation}
where
\begin{equation}\label{rho1}
\rho=\alpha_1 \rho_1 + \alpha_2 \rho_2
\end{equation}

Applying perturbation to Eqn. \ref{NS2}, and neglecting the convection term, we have
\begin{equation}\label{NS3}
\rho \frac{\partial   \mathbf u^{\prime}} {\partial t}+  \nabla p^{\prime}=0
\end{equation}

Combining Eqns. \ref{alpha4} and \ref{NS3}, we obtain the following wave equation
\begin{equation}\label{phom}
\frac{\partial^2 p^{\prime}} {\partial t^2}= \frac{K}{\rho} \nabla^2 p^{\prime}
\end{equation}

Therefore, the speed of sound in the two-phase homogenous flow could be written as
\begin{equation}
\frac{1} {c_{\rm hom}^2}= (\alpha_1 \rho_1 + \alpha_2 \rho_2)\left(\frac{\alpha_1}{\rho_1 c_1^2}+ \frac{\alpha_2}{\rho_2 c_2^2} \right)
\end{equation}

\subsection{Speed of sound in separated flow}
In separated flow, we already mentioned the isobaric condition at each cross section of the pipe. Therefore, the subscript for pressure could also be omitted in this case.

The analysis of continuity equation is the same as in homogeneous flow. Therefore Eqn. \ref{alpha4} is also applied in this case. The major difference comes from the momentum equation as in Eqn. \ref{NS4}.

%The continuity equation for separated flow is the same as Eqn. \ref{alpha1}. And we may derive Eqn. \ref{alpha4} in the same way.

%Consider the Navier-Stokes equation for the separated flow in the longitudinal direction,
%\begin{equation}\label{NS4}
% \rho_k \left( \frac{\partial   u_k} {\partial t}+   u_k \frac{\partial u_k}{\partial x } \right )=- \frac{\partial p}{\partial x}
%\end{equation}
%It is assumed that the flow is inviscid and there is no interfacial shear stress. The jump condition of momentum is written down as Eqn. \ref{pk} if the surface tension is neglected.

Applying perturbation to Eqn. \ref{NS4}, we obtain
\begin{equation}\label{NS5}
 \rho_k  \frac{\partial   u_k^{\prime}} {\partial t} =- \frac{\partial p^{\prime}}{\partial x}
\end{equation}

Combining Eqn. \ref{NS5} for both phases,
\begin{equation}\label{NS6}
 \frac{\partial   u^{\prime}} {\partial t} =-\frac{1}{\rho} \frac{\partial p^{\prime}}{\partial x}
\end{equation}
where
\begin{equation}\label{rho2}
 \frac{1} {\rho}  =  \frac{\alpha_1} {\rho_1} + \frac{\alpha_2} {\rho_2}
\end{equation}

Combining Eqns. \ref{alpha4} and \ref{NS6}, we obtain the speed of sound in inviscid separated flow as,
\begin{equation}
\frac{1} {c_{\rm sep}^2} \left( \frac{\alpha_1}{ \rho_1} + \frac{\alpha_2}{ \rho_2} \right)=  \frac{\alpha_1}{\rho_1 c_1^2}+ \frac{\alpha_2}{\rho_2 c_2^2}
\end{equation}
where $c_k$ is specified in Eqn. \ref{EOSp1}.

\bibliographystyle{elsarticle-harv}
\bibliography{refs}

\begin{thebibliography}{35}
\expandafter\ifx\csname natexlab\endcsname\relax\def\natexlab#1{#1}\fi
\providecommand{\url}[1]{\texttt{#1}}
\providecommand{\href}[2]{#2}
\providecommand{\path}[1]{#1}
\providecommand{\DOIprefix}{doi:}
\providecommand{\ArXivprefix}{arXiv:}
\providecommand{\URLprefix}{URL: }
\providecommand{\Pubmedprefix}{pmid:}
\providecommand{\doi}[1]{\href{http://dx.doi.org/#1}{\path{#1}}}
\providecommand{\Pubmed}[1]{\href{pmid:#1}{\path{#1}}}
\providecommand{\bibinfo}[2]{#2}
\ifx\xfnm\relax \def\xfnm[#1]{\unskip,\space#1}\fi
%Type = Article
\bibitem[{Ardron and Duffey(1978)}]{Ardron1978}
\bibinfo{author}{Ardron, K.H.}, \bibinfo{author}{Duffey, R.B.},
  \bibinfo{year}{1978}.
\newblock \bibinfo{title}{Acoustic wave propagation in a flowing liquid-vapour
  mixture}.
\newblock \bibinfo{journal}{International Journal of Multiphase Flow}
  \bibinfo{volume}{4}, \bibinfo{pages}{303--322}.
%Type = Article
\bibitem[{Baer and Nunziato(1986)}]{Baer1986}
\bibinfo{author}{Baer, M.R.}, \bibinfo{author}{Nunziato, J.W.},
  \bibinfo{year}{1986}.
\newblock \bibinfo{title}{A two-phase mixture theory for the
  deflagration-to-detonation transition (ddt) in reactive granular materials}.
\newblock \bibinfo{journal}{International Journal of Multiphase Flow}
  \bibinfo{volume}{12}, \bibinfo{pages}{861--889}.
%Type = Article
\bibitem[{Bai and Deng(2017)}]{Bai2017}
\bibinfo{author}{Bai, X.}, \bibinfo{author}{Deng, X.}, \bibinfo{year}{2017}.
\newblock \bibinfo{title}{A sharp interface method for compressible multi-phase
  flows based on the cut cell and ghost fluid methods}.
\newblock \bibinfo{journal}{Advances in Applied Mathematics and Mechanics}
  \bibinfo{volume}{9}, \bibinfo{pages}{1052--1075}.
%Type = Inproceedings
\bibitem[{Beuthe(1997)}]{Beuthe1997}
\bibinfo{author}{Beuthe, T.G.}, \bibinfo{year}{1997}.
\newblock \bibinfo{title}{Review of two-phase water hammer}, in:
  \bibinfo{booktitle}{Proceedings of the 18th Canadian Nuclear Society
  Conference}, \bibinfo{address}{Toronto, Canada}.
%Type = Book
\bibitem[{Brennen(2005)}]{Brennen2005}
\bibinfo{author}{Brennen, C.E.}, \bibinfo{year}{2005}.
\newblock \bibinfo{title}{Fundamentals of multiphase flows}.
\newblock \bibinfo{publisher}{Cambridge University Press}.
%Type = Misc
\bibitem[{Calvert(2000)}]{Calvert2000}
\bibinfo{author}{Calvert, J.B.}, \bibinfo{year}{2000}.
\newblock \bibinfo{title}{Water hammer}.
\newblock
  \bibinfo{howpublished}{\url{https://mysite.du.edu/~jcalvert/tech/fluids/waterham.htm}}.
\newblock \bibinfo{note}{[Online; accessed 03-March-2017]}.
%Type = Article
\bibitem[{Chang and Liou(2007)}]{Chang2007}
\bibinfo{author}{Chang, C.}, \bibinfo{author}{Liou, M.}, \bibinfo{year}{2007}.
\newblock \bibinfo{title}{A robust and accurate approach to computing
  compressible multiphase flow: Stratified flow model and ausm${^+}$-up
  scheme}.
\newblock \bibinfo{journal}{Journal of Computational Physics}
  \bibinfo{volume}{225}, \bibinfo{pages}{840--873}.
%Type = Techreport
\bibitem[{Cheng et~al.(1983)Cheng, Drew and Lahey}]{Cheng1983}
\bibinfo{author}{Cheng, L.Y.}, \bibinfo{author}{Drew, D.A.},
  \bibinfo{author}{Lahey, R.T.}, \bibinfo{year}{1983}.
\newblock \bibinfo{title}{An analysis of wave dispersion, sonic velocity, and
  critical flow in two-phase mixtures}.
\newblock \bibinfo{type}{Technical Report} \bibinfo{number}{NUREG/CR-3372}.
  Rensselaer Polytechnic Institute.
\newblock \URLprefix \url{https://ntrl.ntis.gov/NTRL/}.
%Type = Article
\bibitem[{Cheng et~al.(1985)Cheng, Drew and Lahey}]{Cheng1985}
\bibinfo{author}{Cheng, L.Y.}, \bibinfo{author}{Drew, D.A.},
  \bibinfo{author}{Lahey, R.T.}, \bibinfo{year}{1985}.
\newblock \bibinfo{title}{An analysis of wave propagation in bubbly
  two-component, two-phase flow}.
\newblock \bibinfo{journal}{Journal of Heat Transfer} \bibinfo{volume}{107},
  \bibinfo{pages}{402--408}.
%Type = Incollection
\bibitem[{Corradini et~al.(2016)Corradini, Zhu, Fan and Jean}]{Corradini2016}
\bibinfo{author}{Corradini, M.L.}, \bibinfo{author}{Zhu, C.},
  \bibinfo{author}{Fan, L.}, \bibinfo{author}{Jean, R.}, \bibinfo{year}{2016}.
\newblock \bibinfo{title}{Multiphase flow}, in: \bibinfo{editor}{Johnson, R.W.}
  (Ed.), \bibinfo{booktitle}{Handbook of Fluid Dynamics}.
  \bibinfo{publisher}{CRC Press}, \bibinfo{address}{Boca Raton}.
  chapter~\bibinfo{chapter}{20}.
%Type = Article
\bibitem[{Costigan and Whalley(1997)}]{Costigan1997}
\bibinfo{author}{Costigan, G.}, \bibinfo{author}{Whalley, P.B.},
  \bibinfo{year}{1997}.
\newblock \bibinfo{title}{Measurements of the speed of sound in air-water
  flows}.
\newblock \bibinfo{journal}{Chemical Engineering Journal} \bibinfo{volume}{66},
  \bibinfo{pages}{131--135}.
%Type = Phdthesis
\bibitem[{Dijk(2005)}]{Dijk2005}
\bibinfo{author}{Dijk, P.v.}, \bibinfo{year}{2005}.
\newblock \bibinfo{title}{Acoustics of Two-Phase Pipe Flows}.
\newblock Ph.D. thesis. University of Twente. \bibinfo{address}{Enschede,
  Netherlands}.
%Type = Book
\bibitem[{Drew and Passman(1999)}]{Drew1999}
\bibinfo{author}{Drew, D.A.}, \bibinfo{author}{Passman, S.L.},
  \bibinfo{year}{1999}.
\newblock \bibinfo{title}{Theory of Multicomponent Fluids}.
\newblock \bibinfo{publisher}{Springer}.
%Type = Article
\bibitem[{{Drui} et~al.(2016){Drui}, {Larat}, {Kokh} and {Massot}}]{Drui2016}
\bibinfo{author}{{Drui}, F.}, \bibinfo{author}{{Larat}, A.},
  \bibinfo{author}{{Kokh}, S.}, \bibinfo{author}{{Massot}, M.},
  \bibinfo{year}{2016}.
\newblock \bibinfo{title}{{A hierarchy of simple hyperbolic two-fluid models
  for bubbly flows}}.
\newblock \bibinfo{journal}{ArXiv e-prints}
  \href{http://arxiv.org/abs/1607.08233}{{\tt arXiv:1607.08233}}.
%Type = Misc
\bibitem[{Ernesto(2006)}]{Ernesto2006}
\bibinfo{author}{Ernesto, G.M.}, \bibinfo{year}{2006}.
\newblock \bibinfo{title}{Conduction heat transfer}.
\newblock
  \bibinfo{howpublished}{\url{http://www.ewp.rpi.edu/hartford/~ernesto/S2006/CHT/}}.
\newblock \bibinfo{note}{[Online; accessed 11-Apr-2017]}.
%Type = Article
\bibitem[{Fl{\aa}tten et~al.(2010)Fl{\aa}tten, Morin and
  Munkejord}]{Flatten2010}
\bibinfo{author}{Fl{\aa}tten, T.}, \bibinfo{author}{Morin, A.},
  \bibinfo{author}{Munkejord, S.T.}, \bibinfo{year}{2010}.
\newblock \bibinfo{title}{Wave propagation in multicomponent flow models}.
\newblock \bibinfo{journal}{SIAM Journal on Applied Mathematics}
  \bibinfo{volume}{70}, \bibinfo{pages}{2861--2882}.
%Type = Article
\bibitem[{Fox et~al.(1995)Fox, Curley and Larson}]{Fox1955}
\bibinfo{author}{Fox, F.E.}, \bibinfo{author}{Curley, S.R.},
  \bibinfo{author}{Larson, G.S.}, \bibinfo{year}{1995}.
\newblock \bibinfo{title}{Phase velocity and absorption measurements in water
  containing air bubbles}.
\newblock \bibinfo{journal}{The Journal of the Acoustical Society of America}
  \bibinfo{volume}{27}, \bibinfo{pages}{534--539}.
%Type = Article
\bibitem[{{Fu} and {Anglart}(2017)}]{Fu2017}
\bibinfo{author}{{Fu}, K.}, \bibinfo{author}{{Anglart}, H.},
  \bibinfo{year}{2017}.
\newblock \bibinfo{title}{{Implementation and validation of two-phase boiling
  flow models in OpenFOAM}}.
\newblock \bibinfo{journal}{ArXiv e-prints}
  \href{http://arxiv.org/abs/1709.01783}{{\tt arXiv:1709.01783}}.
%Type = Misc
\bibitem[{Hall(2015)}]{Hall2015}
\bibinfo{author}{Hall, N.}, \bibinfo{year}{2015}.
\newblock \bibinfo{title}{Mach number: Role in compressible flows}.
\newblock
  \bibinfo{howpublished}{\url{https://www.grc.nasa.gov/www/k-12/airplane/machrole.html}}.
\newblock \bibinfo{note}{[Online; accessed 03-March-2017]}.
%Type = Book
\bibitem[{Hdaneshyar(1976)}]{Hdaneshyar1976}
\bibinfo{author}{Hdaneshyar, H.}, \bibinfo{year}{1976}.
\newblock \bibinfo{title}{One-dimensional compressible flow}.
\newblock \bibinfo{edition}{1} ed., \bibinfo{publisher}{Pergamon press}.
%Type = Book
\bibitem[{Ishii and Hibiki(2011)}]{Ishii2011}
\bibinfo{author}{Ishii, M.}, \bibinfo{author}{Hibiki, T.},
  \bibinfo{year}{2011}.
\newblock \bibinfo{title}{Thermo-fluid dynamics of two-phase flow}.
\newblock \bibinfo{edition}{2} ed., \bibinfo{publisher}{Springer}.
%Type = Techreport
\bibitem[{Karplus(1958)}]{Karplus1958}
\bibinfo{author}{Karplus, H.B.}, \bibinfo{year}{1958}.
\newblock \bibinfo{title}{The velocity of sound in a liquid containing gas
  bubbles}.
\newblock \bibinfo{type}{Technical Report}. U.S. Atomic Energy Commission.
%Type = Article
\bibitem[{Kieffer(1977)}]{Kieffer1977}
\bibinfo{author}{Kieffer, S.W.}, \bibinfo{year}{1977}.
\newblock \bibinfo{title}{Sound speed in liquid-gas mixtures: Water-air and
  water-steam}.
\newblock \bibinfo{journal}{Journal of Geophysical Research}
  \bibinfo{volume}{82}, \bibinfo{pages}{2895--3118}.
%Type = Article
\bibitem[{Mecredy and Hamilton(1972)}]{Mecredy1972}
\bibinfo{author}{Mecredy, R.C.}, \bibinfo{author}{Hamilton, L.J.},
  \bibinfo{year}{1972}.
\newblock \bibinfo{title}{The effects of nonequilibrium heat, mass and momentum
  transfer on two-phase sound speed}.
\newblock \bibinfo{journal}{International Journal of Heat and Mass Transfer}
  \bibinfo{volume}{15}, \bibinfo{pages}{61--72}.
%Type = Article
\bibitem[{Murrone and Guillard(2005)}]{Murrone2005}
\bibinfo{author}{Murrone, A.}, \bibinfo{author}{Guillard, H.},
  \bibinfo{year}{2005}.
\newblock \bibinfo{title}{A five equation reduced model for compressible two
  phase flow problems}.
\newblock \bibinfo{journal}{Journal of Computational Physics}
  \bibinfo{volume}{202}, \bibinfo{pages}{664--698}.
%Type = Article
\bibitem[{Ranjan et~al.(2011)Ranjan, Oakley and Bonazza}]{Ranjan2011}
\bibinfo{author}{Ranjan, D.}, \bibinfo{author}{Oakley, J.},
  \bibinfo{author}{Bonazza, R.}, \bibinfo{year}{2011}.
\newblock \bibinfo{title}{Shock-bubble interactions}.
\newblock \bibinfo{journal}{Annual Review of Fluid Mechanics}
  \bibinfo{volume}{43}, \bibinfo{pages}{117--140}.
%Type = Article
\bibitem[{Ruggles et~al.(1988)Ruggles, Lahey, Drew and Scarton}]{Ruggles1988}
\bibinfo{author}{Ruggles, A.E.}, \bibinfo{author}{Lahey, R.T.},
  \bibinfo{author}{Drew, D.A.}, \bibinfo{author}{Scarton, H.A.},
  \bibinfo{year}{1988}.
\newblock \bibinfo{title}{An investigation of the propagation of pressure
  perturbations in bubbly air/water flows}.
\newblock \bibinfo{journal}{Journal of Heat Transfer} \bibinfo{volume}{110},
  \bibinfo{pages}{494--499}.
%Type = Article
\bibitem[{Ruggles et~al.(1989)Ruggles, Lahey, Drew and Scarton}]{Ruggles1989}
\bibinfo{author}{Ruggles, A.E.}, \bibinfo{author}{Lahey, R.T.},
  \bibinfo{author}{Drew, D.A.}, \bibinfo{author}{Scarton, H.A.},
  \bibinfo{year}{1989}.
\newblock \bibinfo{title}{The relationship between standing waves, pressure
  pulse propagation, and critical flow rate in two-phase mixtures}.
\newblock \bibinfo{journal}{Journal of Heat Transfer} \bibinfo{volume}{111},
  \bibinfo{pages}{467--473}.
%Type = Article
\bibitem[{Saurel et~al.(2009)Saurel, Petitpas and Berry}]{Saurel2009}
\bibinfo{author}{Saurel, R.}, \bibinfo{author}{Petitpas, F.},
  \bibinfo{author}{Berry, R.A.}, \bibinfo{year}{2009}.
\newblock \bibinfo{title}{Simple and efficient relaxation methods for
  interfaces separating compressible fluids, cavitating flows and shocks in
  multiphase mixtures}.
\newblock \bibinfo{journal}{Journal of Computational Physics}
  \bibinfo{volume}{228}, \bibinfo{pages}{1678--1712}.
%Type = Article
\bibitem[{Shyue(1998)}]{Shyue1998}
\bibinfo{author}{Shyue, K.}, \bibinfo{year}{1998}.
\newblock \bibinfo{title}{An efficient shock-capturing algorithm for
  compressible multicomponent problems}.
\newblock \bibinfo{journal}{Journal of Computational Physics}
  \bibinfo{volume}{142}, \bibinfo{pages}{208--242}.
%Type = Misc
\bibitem[{Shyue(2014)}]{Shyue2014}
\bibinfo{author}{Shyue, K.}, \bibinfo{year}{2014}.
\newblock \bibinfo{title}{Recent advances in numerical methods for compressible
  two-phase flow with heat \& mass transfers}.
\newblock
  \bibinfo{howpublished}{\url{http://www.math.ntu.edu.tw/~shyue/mytalks/kmshyue_twcfd2014.pdf}}.
\newblock \bibinfo{note}{[Online; accessed 31-August-2017]}.
%Type = Article
\bibitem[{Simon et~al.(2016)Simon, Martinez-Molina and
  Fortes-Patella}]{Simon2016}
\bibinfo{author}{Simon, A.}, \bibinfo{author}{Martinez-Molina, J.},
  \bibinfo{author}{Fortes-Patella, R.}, \bibinfo{year}{2016}.
\newblock \bibinfo{title}{A new process to estimate the speed of sound using
  three-sensor method}.
\newblock \bibinfo{journal}{Experiments in Fluids} \bibinfo{volume}{57},
  \bibinfo{pages}{10}.
%Type = Book
\bibitem[{Toro(2009)}]{Toro2009}
\bibinfo{author}{Toro, E.F.}, \bibinfo{year}{2009}.
\newblock \bibinfo{title}{Riemann Solvers and Numerical Methods for Fluid
  Dynamics}.
\newblock \bibinfo{edition}{3} ed., \bibinfo{publisher}{Springer}.
%Type = Inproceedings
\bibitem[{Wijngaarden(1976)}]{Wijngaarden1976}
\bibinfo{author}{Wijngaarden, L.}, \bibinfo{year}{1976}.
\newblock \bibinfo{title}{Some problems in the formulation of the equations for
  gas/liquid flows}, in: \bibinfo{booktitle}{14th IUTAM Congress on Theoretical
  and Applied Mechanics}, \bibinfo{address}{Delft, the Netherlands}. pp.
  \bibinfo{pages}{249--260}.
%Type = Article
\bibitem[{Zein et~al.(2010)Zein, Hantke and Warnecke}]{Zein2010}
\bibinfo{author}{Zein, A.}, \bibinfo{author}{Hantke, M.},
  \bibinfo{author}{Warnecke, G.}, \bibinfo{year}{2010}.
\newblock \bibinfo{title}{Modeling phase transition for compressible two-phase
  flows applied to metastable liquids}.
\newblock \bibinfo{journal}{Journal of Computational Physics}
  \bibinfo{volume}{229}, \bibinfo{pages}{2964--2998}.

\end{thebibliography}

\end{document}